\documentclass[fleqn,usenatbib]{mnras}

\usepackage{newtxtext,newtxmath}
\usepackage{subcaption}

\usepackage[T1]{fontenc}

\DeclareRobustCommand{\VAN}[3]{#2}
\let\VANthebibliography\thebibliography
\def\thebibliography{\DeclareRobustCommand{\VAN}[3]{##3}\VANthebibliography}

\usepackage{graphicx}	
\usepackage{amsmath}	
\usepackage{xcolor}		
\usepackage[normalem]{ulem}	
\usepackage{microtype}		
\title[Cosmic Variance at $z \approx 6$]{The Impact of Cosmic Variance and Satellites on JWST Clustering Measurements at redshift $\approx 6$}

\author[J. Huang et al.]{
  Jiamu Huang$^{1}$\thanks{E-mail: jiamu\_huang@ucsb.edu},
  Elia Pizzati$^{2}$\thanks{NHFP Einstein Fellow},
  Joseph Hennawi$^{1, 3}$,
  Joop Schaye$^{3}$,
  Matthieu Schaller$^{3,4}$,
    \newauthor
  ~Benjamin Snyder$^{1}$,
  Yi Kang$^{1}$
  \\
  $^{1}$Department of Physics, University of California, Santa Barbara, CA 93106, USA \\
  $^{2}$ Center for Astrophysics | Harvard \& Smithsonian, 60 Garden St., Cambridge, MA 02138, USA \\
  $^{3}$Leiden Observatory, Leiden University, NL-2300 RA Leiden, The Netherlands \\
  $^{4}$Lorentz Institute for Theoretical Physics, Leiden University, PO Box 9506, NL\-2300 RA Leiden, The Netherlands\\
  }

\date{Accepted XXX. Received YYY; in original form ZZZ}

\pubyear{2026}

\begin{document}

\font\sevenrm=cmr7
\def\oiii{[O\,\textsc{iii}]}
\def\OIII{[O~{\sevenrm III}]}
\def\FeII{Fe~{\sevenrm II}}
\def\FeIIf{[Fe~{\sevenrm II}]}
\def\SIII{[S~{\sevenrm III}]}
\def\HeI{He~{\sevenrm I}}
\def\HeII{He~{\sevenrm II}}
\def\NeV{[Ne~{\sevenrm V}]}
\def\OIV{[O~{\sevenrm IV}]}

\def\mpch{~\mathrm{cMpc}~h^{-1}}
\def\rp{r_{p}}
\def\msun{{\rm {M_{\odot}}}}
\def\mminq{M^{\rm QSO}_{h, \rm min}}
\def\mming{M^{\rm \oiii}_{h, \rm min}}
\def\mmincen{M^{\rm cen,\,\oiii}_{h, \rm min}}
\def\mminsat{M^{\rm sat,\,\oiii}_{h, \rm min}}

\label{firstpage}
\pagerange{\pageref{firstpage}--\pageref{lastpage}}
\maketitle

\begin{abstract}
We present a framework for inferring the dark matter halo masses of quasars and \OIII-emitting galaxies from JWST/NIRCam Wide Field Slitless Spectroscopy (WFSS) clustering measurements at $z \approx 6$. Using the FLAMINGO-10k N-body simulation, we construct mock realizations of quasar and galaxy catalogs that incorporate realistic selection functions, spatial coverage, and sensitivity limits matched to the ASPIRE survey. These mocks enable accurate measurements of the quasar--galaxy cross-correlation and galaxy auto-correlation functions, with covariance matrices derived from 1000 realizations that capture both cosmic variance and bin-to-bin correlations. We employ Bayesian inference to fit the correlation functions and infer the minimum halo masses for quasars and galaxies. Our results demonstrate that Poisson pair-count uncertainties, commonly adopted in high-redshift clustering studies, significantly underestimate the true measurement errors. The dominant missing component is cosmic variance: even the diagonal of the full covariance matrix exceeds the Poisson expectation, with off-diagonal bin-to-bin correlations contributing a smaller additional correction. In particular, 1) the commonly used Poisson error on the correlation functions underestimates the true uncertainty by a factor of $\approx 3$; 2) the uncertainties on the inferred minimum halo masses are underestimated by a factor of $\approx 1.5$--$3$ when adopting Poisson errors instead of the full covariance matrix; 3) the inferred QSO halo mass is robust to whether central and satellite \oiii-emitters share a common mass threshold. Our framework provides a more complete error budget for JWST/WFSS clustering analyses, enabling robust constraints on the host halo masses and duty cycles of high-redshift quasars and emission-line galaxies.
\end{abstract}




\section{Introduction}
\label{sec:introduction}

The detection of quasars at high redshift hosting supermassive black holes (SMBHs) with masses of $\sim 10^9~\msun$ \citep{Mortlock2011, Matsuoka2019, Yang2020, Wang2021, Fan2023} poses significant challenges to our understanding of black hole formation and growth in the early Universe. To observationally constrain the growth scenarios of early SMBHs, it is essential to determine not only how massive these black holes are when active, but also what fraction of their growth occurs during the UV-bright phase that we observe as quasars. Quasar clustering provides a direct way to quantify this through measurements of the quasar duty cycle $f_{\rm duty} \equiv n_{\rm quasar}/n_{\rm halo}$, the ratio of the observed quasar number density to the number density of their host dark matter halos \citep{Efstathiou1988, Cole1989, HaimanHui2001, Martini2001}. Since more massive dark matter halos exhibit stronger clustering due to their higher bias \citep{Kaiser1984, MoWhite2002}, we can statistically infer the typical host halo mass of quasars from the observed clustering strength, and thereby determine the duty cycle from the abundance of halos above that mass threshold. Under the assumption that each halo undergoes one quasar phase, the duty cycle is equivalent to the quasar lifetime as a fraction of the Hubble time, $f_{\rm duty} \approx t_{\rm Q}/t_{\rm H}$ \citep{Efstathiou1988, Cole1989, HaimanHui2001, Martini2001}.

Clustering analysis through the two-point correlation function (2PCF) has become a powerful tool to probe the environments and host halo masses of quasars \citep{Osmer1981, Shen2007}. At redshifts $1 \lesssim z \lesssim 3$, quasar clustering measurements indicate host halo masses of order of $10^{12}~\msun$, increasing to $\sim 10^{13}~\msun$ at $z \approx 4$ \citep{Porciani2004, Croom2005, Coil2007, Pizzati2024a,GinerMascarell2025}. 

However, high-redshift measurements of quasar auto-correlation functions become challenging due to the low spatial density of luminous quasars at $z \gtrsim 6$ \citep{Wang2019, Schindler2023}. Recently, \citet{Arita2023} measured the quasar auto-correlation at $z \approx 6$ by targeting the fainter, more abundant quasar population ($M_{1450} \approx -25$) from the SHELLQs survey, although the resulting constraints on the host halo mass carry large statistical uncertainties due to the limited sample size, motivating the use of cross-correlation measurements with more abundant tracers.  
For the rarer, more luminous quasars, cross-correlation measurements between quasars and galaxies offer a complementary approach for determining host halo masses at high redshift \citep{GarciaVergara2017, GarciaVergara2019, GarciaVergara2022, Eilers2024, Pizzati2024b, Schindler2026}.

Before \textit{JWST}, empirical studies of quasar environments at $z > 5$ yielded mixed results: narrow band searches for Ly$\alpha$ emitters (LAEs) and Lyman break galaxies (LBGs) sometimes reported galaxy excesses around high-$z$ quasars \citep[e.g.,][]{Kim2009, Morselli2014}, whereas others found no significant overdensities \citep{Banados2013, Simpson2014}. These mixed conclusions likely reflected observational limitations, including small fields of view, heterogeneous depth, redshift offsets between narrow band filters and quasar systemic redshifts, and the heterogeneity of the LAE/LBG tracer populations themselves, whose host halo masses are poorly constrained but directly affect the interpretation of the cross-correlation signal \citep{Overzier2016}. More recently, \citet{Arita2026} studied four low-luminosity quasars ($-24 < M_{1450} < -22$) at $z \approx 6.2$ using Subaru/HSC narrowband imaging of Ly$\alpha$ emitters, finding that only one of four fields shows an overdensity ($\delta_{\rm LAE} = 3.77 \pm 0.97$) while the remaining three are consistent with the field, further illustrating the diversity of quasar environments inferred from narrowband studies.

Recent observations with \textit{JWST}/NIRCam using the Wide Field Slitless Spectroscopy (WFSS) mode have opened new opportunities to probe quasar environments with higher sensitivity and larger spatial coverage. Several ongoing programs are now using WFSS to search for line-emitting galaxies, including \OIII\ and H$\alpha$ emitters, in quasar fields. These studies have revealed an abundance of \OIII-emitting galaxies around some high-redshift quasars with $z \gtrsim 6$ \citep{Wang2023, Kashino2023, Eilers2024, Champagne2025}, suggesting that luminous quasars may reside in significant overdensities. However, \citet{Schindler2026} measured the quasar--\OIII\ emitter cross-correlation for two quasars at $z \simeq 7.3$ and found a weaker clustering signal, indicating that quasar environments may vary from object to object.
The growing sample of emission-line galaxy tracers has enabled more precise clustering analyses: the combination of \OIII-emitter auto-correlation and quasar--\OIII-emitter cross-correlation measurements has yielded the first joint constraints on the dark matter halo masses hosting galaxies and quasars at $z > 6$ \citep{Eilers2024, Pizzati2024b, Huang2026}, with achievable measurement precision broadly consistent with pre-launch forecasts of \textit{JWST}/WFSS clustering uncertainties \citep{Endsley2020}. 

Beyond quasar--galaxy cross-correlations, \textit{JWST}/NIRCam WFSS has enabled galaxy auto-correlation measurements at high redshift. \citet{Shuntov2025} measured the angular auto-correlation of H$\alpha$ emitters at $3.7 < z < 5.1$ (CONGRESS, F356W) and $4.9 < z < 6.7$ (FRESCO, F444W; \citealt{Oesch2023}), and \OIII\ emitters at $6.8 < z < 9.0$ (FRESCO, F444W) in the GOODS fields. Through HOD modeling of the clustering and UV luminosity function, they inferred characteristic halo masses decreasing from $\log(M_h/\msun) = 11.46$ at $\langle z \rangle = 4.3$ to $10.99$ at $\langle z \rangle = 7.3$. \citet{Lin2025_lf} measured the auto-correlation of H$\alpha$ emitters at $z = 4$--$6$ using the same CONGRESS+FRESCO data, finding correlation lengths of $r_0 = 4.61^{+1.00}_{-0.68}~\mpch$ at $z = 4$--$5$ and $r_0 = 6.23^{+1.68}_{-1.13}~\mpch$ at $z = 5$--$6$, corresponding to host halo masses of $\log(M_h/\msun) \approx 11.0$--$11.2$. \citet{Lin2025} further measured the AGN--H$\alpha$ emitter cross-correlation at $3.9 < z < 6$ and found a comparable clustering amplitude to the HAE auto-correlation, indicating that low-luminosity AGNs share similar bias and host halo masses ($\log(M_h/\msun) \approx 11.0$--$11.2$) with HAEs. This places them in a population distinct from the more biased, more massive luminous quasars.

From the theoretical perspective, galaxy clustering at high redshift breaks a key degeneracy in the UV luminosity function: the same observed number counts can be produced either by bright galaxies in rare, massive halos (high duty cycle) or by occasional bright phases of more common, lower-mass halos (low duty cycle). Since these two scenarios predict different clustering amplitudes, measuring the galaxy auto-correlation function constrains the typical host halo mass and duty cycle of the bright UVLF tracers \citep{Munoz2023}. \citet{Weibel2025} recently applied this approach at $z \approx 10$, using 34 independent NIRCam sightlines from PANORAMIC to measure the cosmic variance of Lyman-break galaxies and, combined with the UV luminosity function, disfavor models with globally enhanced star-formation efficiency.

A critical but underappreciated issue in all existing high-$z$ clustering analyses is the treatment of measurement uncertainties. It is common practice in high-redshift quasar clustering studies to estimate the uncertainty of the correlation function using Poisson errors on the pair counts \citep{GarciaVergara2017, GarciaVergara2019, Eilers2024, Schindler2026}. This approach assumes that the covariance matrix is diagonal and ignores two key sources of correlated uncertainty. First, different galaxy pairs are not statistically independent because the same galaxies contribute to multiple pairs, so any large-scale fluctuation in the galaxy density (for example, a locally overdense region) boosts the number of pairs across a wide range of separations, producing positive correlations between radial bins. Second, cosmic variance, i.e.\ the field-to-field variation in the large-scale density field, introduces coherent fluctuations 
that affect all bins simultaneously. The Poisson approximation therefore underestimates not only the diagonal error bars but also completely neglects the off-diagonal covariances between radial bins, so that the true error budget is substantially larger and more correlated than implied by Poisson statistics alone. This effect has recently been quantified using the FLAMINGO suite by \citet{Lim2025}, who showed that for JWST-like (100~cMpc)$^3$ survey volumes at $z \simeq 6$, the field-to-field variance in massive-halo number counts is 2--3 times the Poisson expectation, with a comparable enhancement in the variance of the most-massive halo per field.

With only a few independent quasar fields per program (e.g., four in EIGER, \citealt{Eilers2024}), internal error estimators such as jackknife or bootstrap resampling cannot provide stable covariance estimates: the resulting covariance matrices are typically noisy and can become poorly conditioned or even singular \citep{Norberg2009}. This effectively forces high-$z$ studies \citep{Eilers2024, Schindler2026}
to adopt Poisson pair-count uncertainties, leaving the reported error bars as underestimates of the true uncertainty \citep{Yang2005, Robertson2010}. In contrast, at low-to-intermediate redshifts ($z \lesssim 3$), the methodology for robust clustering inference is well established. Large galaxy surveys such as the Sloan Digital Sky Survey \citep[SDSS;][]{Zehavi2011, Zheng2007, Guo2015}, the Baryon Oscillation Spectroscopic Survey \citep[BOSS;][]{Manera2013, White2014, Kitaura2016}, the Dark Energy Survey \citep[DES;][]{Friedrich2021}, and most recently the Dark Energy Spectroscopic Instrument \citep[DESI;][]{ForeroSanchez2025} have invested heavily in constructing hundreds to thousands of mock galaxy catalogs from N-body simulations or fast approximate methods \citep[see][for a systematic comparison]{Lippich2019} to estimate covariance matrices for correlation function measurements. These covariance matrices are then used directly in HOD fitting of the volume-averaged correlation function $w_p(r_p)$ to infer halo masses and the galaxy--halo connection \citep{Sinha2018, vandenBosch2013}.

Empirical covariance estimators such as jackknife or bootstrap resampling require many independent sub-volumes to converge \citep{Norberg2009}, and the small number of independent ASPIRE fields makes such observation-based estimators unreliable in our regime. We therefore compute the covariance from forward-modeled mock catalogs, which capture both cosmic variance and correlated bin-to-bin uncertainties. These mock ensembles provide two key ingredients: (i) mock covariance matrices that capture cosmic variance and correlated bin-to-bin uncertainties, and (ii) simulation-based clustering models that correctly account for non-linear structure growth and scale-dependent bias at small separations. The importance of full covariance modeling has been demonstrated repeatedly. \citet{Norberg2009} showed that internal error estimators such as jackknife and bootstrap are biased relative to ensemble-based covariance from independent mock realizations, with neither method reliably reproducing the off-diagonal covariance structure on scales of $1$--$25~h^{-1}$~Mpc \citep{MohammadPercival2022}. The practical requirements for the number of mock realizations needed to obtain an unbiased inverse covariance matrix have been quantified by \citet{Hartlap2007}, \citet{DodelsonSchneider2013}, and \citet{Percival2014}, who showed that using too few mocks inflates parameter confidence intervals and that $\mathcal{O}(10^3)$ realizations are typically required for stable estimates \citep[see also][]{SelentinHeavens2016}. Moreover, within the framework of the halo occupation distribution \citep[HOD;][]{BerlindWeinberg2002, Zheng2005}, the covariance matrix itself depends on the assumed halo mass parameters: more massive, more biased tracers produce stronger cosmic variance and more prominent off-diagonal correlations. As we show in Section~\ref{sec:cov_vs_poisson}, ignoring this mass dependence of the covariance, or replacing the full covariance with diagonal Poisson errors, underestimates the true uncertainty on inferred halo masses by factors of $\approx 1.5$--$3\times$.

The need for proper covariance modeling is even more acute at high redshift. The narrow pencil-beam fields of \textit{JWST}/WFSS ($\approx 2' \times 2'$ each) make cosmic variance the dominant source of uncertainty \citep{Yang2005, TrentiStiavelli2008, Moster2011}, and the highly biased tracers at $z > 5$ further amplify the effect. 
In our companion paper \citep{Huang2026}, we apply these mass-dependent covariance matrices to the ASPIRE observations and show that    
adopting Poisson errors underestimates the $1\sigma$ uncertainty on the inferred halo masses by about $0.7$~dex for quasar minimum halo mass ($\mminq$) and $0.5$~dex for galaxy host minimum mass ($\mming$).

In addition to getting the error budget correct, simulations are also required to physically interpret the measured clustering amplitude in terms of halo masses. In previous high-redshift studies, halo mass estimates are commonly obtained by fitting a power-law model to the large-scale clustering signal, converting the fitted amplitude into an effective bias, and then associating this bias with a characteristic halo mass using analytical prescriptions calibrated on linear theory \citep[e.g.,][]{Arita2023}. This approach is not appropriate
for \textit{JWST} WFSS measurements of small pencil-beam fields,
which probe small scales ($r_p \sim 0.1$--$5~\mpch$) and highly biased tracers such as quasars at high redshift. At these redshifts and separations, the halo bias becomes strongly non-linear and scale-dependent, and standard analytical bias models do not capture these effects. In the quasi-linear regime near the one-halo to two-halo transition, analytical models can significantly underpredict the clustering amplitude if these non-linearities are not treated correctly \citep[see discussions in][]{Jose2016, Shuntov2025}. Recent work has also shown that fitting a rigid power-law correlation function (as in e.g.\ \citealt{Arita2023, Eilers2024}) can lead to biased halo mass estimates that reflect the assumed power-law shape rather than the underlying halo population \citep[demonstrated on large scales in Appendix Fig.~C1 of][]{Pizzati2024b}. In contrast, our simulation-based clustering model links the observed signal directly to the spatial distribution of halos in the simulation, providing a physically grounded inference of the masses of quasar-host halos.

This work addresses the gap between large-area clustering surveys, where mock covariance methodology is well established, and the small-area, pencil-beam JWST/WFSS regime where cosmic variance dominates the error budget, by constructing mock covariance matrices from the FLAMINGO-10k (F10k) cosmological simulation for JWST-style clustering inference. We employ the F10k dark-matter-only simulation (Schaller et al., in prep.; see also \citealt{Pizzati2024b}), a $(2.8~\mathrm{cGpc})^3$ volume containing $10080^3$ CDM particles from the FLAMINGO suite \citep{Schaye2023, Kugel2023}, to generate realistic mock catalogs of \OIII-emitters around $z \sim 6.6$ quasars. Each mock realization replicates the observational properties of the \textit{JWST} ASPIRE survey, incorporating the field-by-field spatial coverage, spectral sensitivity, and flux limits through the same selection function applied to the real data. From 1000 such realizations, we construct the full covariance matrices for the \OIII-emitter auto-correlation function $\chi_{\rm GG}(r_p)$ and the quasar--\OIII-emitter cross-correlation function $\chi_{\rm QG}(r_p)$. Crucially, these covariance matrices are computed on a two-dimensional grid of minimum halo masses ($\mminq$, $\mming$), capturing the mass dependence of the covariance structure that arises from the bias dependence of cosmic variance.

Our approach differs from prior high-$z$ clustering studies in three specific aspects. First, we use mock covariance matrices based on F10k that capture cosmic variance and off-diagonal bin-to-bin correlations, rather than Poisson pair-count errors. Second, we employ simulation-based clustering models that accurately reproduce the non-linear, scale-dependent halo bias at small separations, rather than analytical power-law prescriptions. Third, we incorporate the full mass dependence of the covariance matrix into the likelihood, so that the error budget is self-consistently tied to the halo mass model. Together, these elements bring the robust statistical methodology of low-$z$ clustering analyses to the high-$z$ regime for the first time.

The paper is structured as follows. In Section~\ref{sec:measuring_cf} we define the volume-averaged correlation function estimators. Section~\ref{sec:selection_function} describes the construction of the selection function, including coverage and sensitivity maps. Section~\ref{sec:cov} presents the mock catalog generation and covariance matrix construction. In Section~\ref{sec:mcmc} we describe the Bayesian 
inference framework and compare constraints from the full covariance to those from Poisson errors. We summarize our conclusions in Section~\ref{sec:conclusion}.

\section{Correlation Function Estimators}
\label{sec:measuring_cf}

We quantify galaxy clustering using the volume-averaged correlation function $\chi$ \citep{Hennawi2006, GarciaVergara2017}, which measures the mean overdensity within cylindrical shells in redshift space. For a shell bounded by projected separations $\rp^{\rm min}$ and $\rp^{\rm max}$, this is defined as:
\begin{equation}
\chi(\rp^{\rm min}, \rp^{\rm max}) = \frac{2}{V} \int_{\rp^{\rm min}}^{\rp^{\rm max}} \int_{0}^{\pi_{\max}} \xi(\rp, \pi) \, 2 \pi \rp \, d\rp \, d\pi,
\label{eq:volavg_corr}
\end{equation}
where $\xi(\rp, \pi)$ is the two-dimensional correlation function, $\pi$ is the comoving line-of-sight separation, and $V$ is the cylindrical shell volume bounded by $\rp^{\rm min}$, $\rp^{\rm max}$, and $\pi_{\max}$.
Following \citet{Huang2026}, we adopt $\pi_{\max} = 7\,\mpch$, corresponding to a line-of-sight velocity of ${\rm d}v = 1037\,{\rm km\,s^{-1}}$ at $z=6.5$. The pair counts entering these estimators are computed with \texttt{Corrfunc} \citep{Sinha2020}, a suite of highly optimized, OPENMP parallel clustering codes.

For the \OIII-emitter auto-correlation, we use the Landy--Szalay estimator \citep{LZ1993}:
\begin{equation}
\chi_{\text{GG}}(\rp^{\rm min}, \rp^{\rm max}, \pi_{\rm max}) = \frac{\langle D_{\rm G} D_{\rm G} \rangle - 2 \langle D_{\rm G} R_{\rm G} \rangle + \langle R_{\rm G} R_{\rm G} \rangle}{\langle R_{\rm G} R_{\rm G} \rangle},
\label{eq:auto}
\end{equation}
where $\langle D_{\rm G} D_{\rm G} \rangle$, $\langle D_{\rm G} R_{\rm G} \rangle$, and $\langle R_{\rm G} R_{\rm G} \rangle$ are the normalized data--data, data--random, and random--random pair counts, respectively. For the quasar--\OIII-emitter cross-correlation, each ASPIRE pointing contains only a single quasar at a fixed location, so no random quasar catalog is generated. We therefore adopt the \citet{Davis1983} estimator:
\begin{equation}
\chi_{\text{QG}}(\rp^{\rm min}, \rp^{\rm max}, \pi_{\rm max}) = \frac{\langle D_{\rm Q} D_{\rm G} \rangle}{\langle D_{\rm Q} R_{\rm G} \rangle} - 1,
\label{eq:cross}
\end{equation}
where $\langle D_{\rm Q} D_{\rm G} \rangle$ and $\langle D_{\rm Q} R_{\rm G} \rangle$ are the normalized quasar--galaxy and quasar--random pair counts. The normalization of all pair counts follows the convention described in \citet{Huang2026}.

\section{Completeness and Selection Function}
\label{sec:selection_function}

Accurate random catalogs for the Landy--Szalay and Davis--Peebles estimators require two products that characterize the ASPIRE selection function: a spectral coverage map and a sensitivity map. We briefly summarize their construction here, with a more detailed discussion provided in \citet{Huang2026}.

\subsection{Coverage Map}
\label{sec:coverage}

The spectral coverage map determines whether an \OIII-emitter at a given sky position and redshift has both \OIII\ $\lambda\lambda$4960,5008 lines falling within the WFSS detector boundaries. For each ASPIRE pointing, we generate a grid of test positions from the F356W direct imaging, map each position through the grism WCS and the wavelength-dependent dispersion solution \citep{Sun2024jwst} to all dithered WFSS frames, and flag the source as covered if the dispersed lines fall on the detector in at least one exposure. The coverage map is evaluated over $z = 5.3$--$7.0$, corresponding to a comoving line-of-sight depth of $\approx 485\,\mpch$, to capture the redshift-dependent spatial coverage.

\subsection{Sensitivity Map}
\label{sec:sensitivity}

The sensitivity map accounts for spatial variations in the detection threshold arising from differences in exposure depth, background level, and detector sensitivity between NIRCam modules A and B. A grid of mock \OIII-emitters with fixed flux ($f_{\rm mock} = 10^{-17}\,\rm erg\,s^{-1}\,cm^{-2}$) and line width (${\rm FWHM} = 200\,{\rm km\,s^{-1}}$) is injected into the observed WFSS data and extracted with \texttt{unfold\_JWST} \citep{Wang2023}. At each grid position, the $5\sigma$ flux limit is defined as $F_{\rm lim} \equiv 5\,f_{\rm mock}/(S/N)_{\rm mock}$.

\subsection{Random Catalog Construction}

\begin{figure*}
\centering
    \begin{subfigure}{\columnwidth}
        \centering
        \includegraphics[width=\columnwidth]{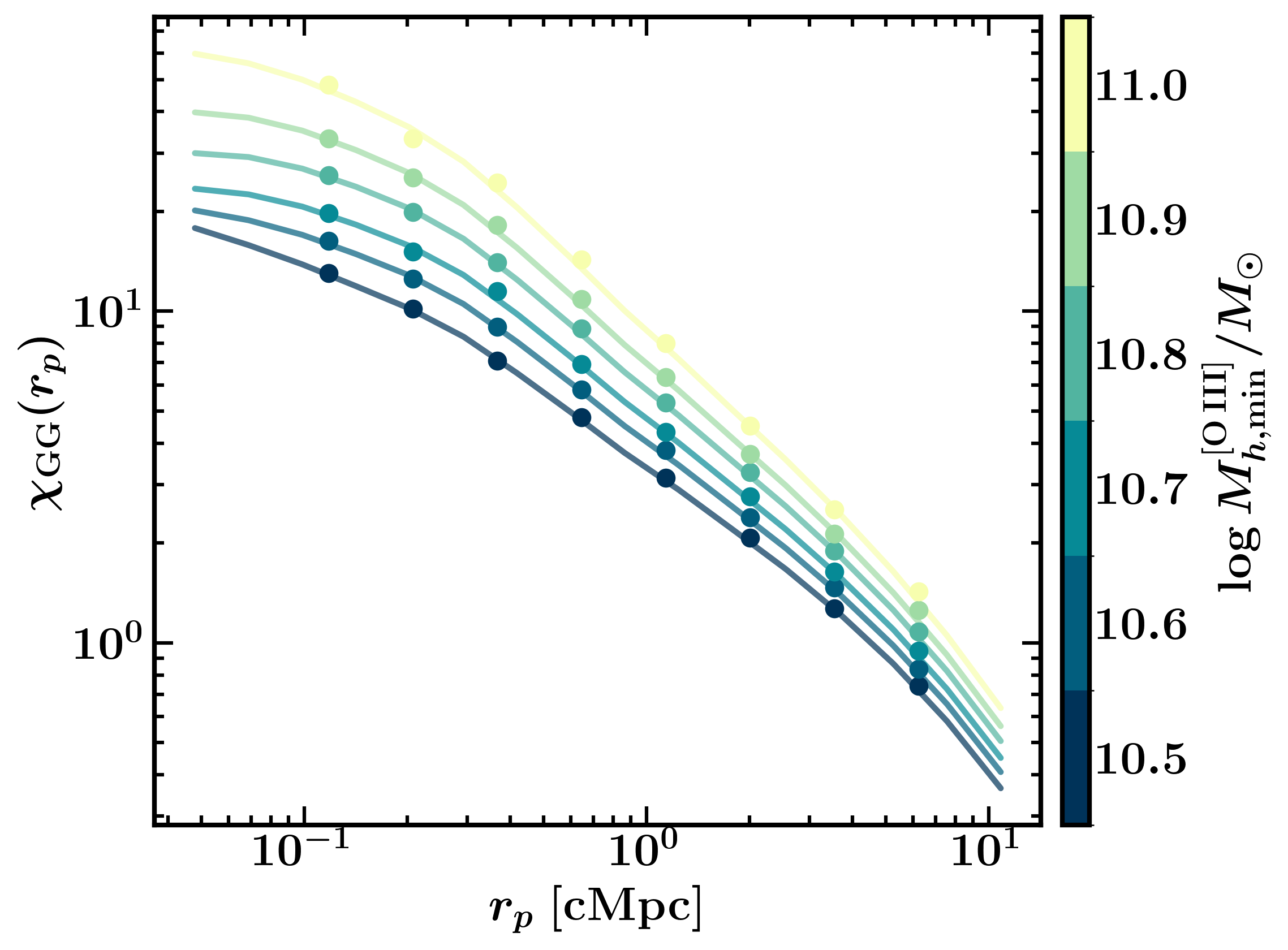}
        \label{fig:gg_corr}
    \end{subfigure}
    \hfill
    \begin{subfigure}{\columnwidth}
        \centering
        \includegraphics[width=\columnwidth]{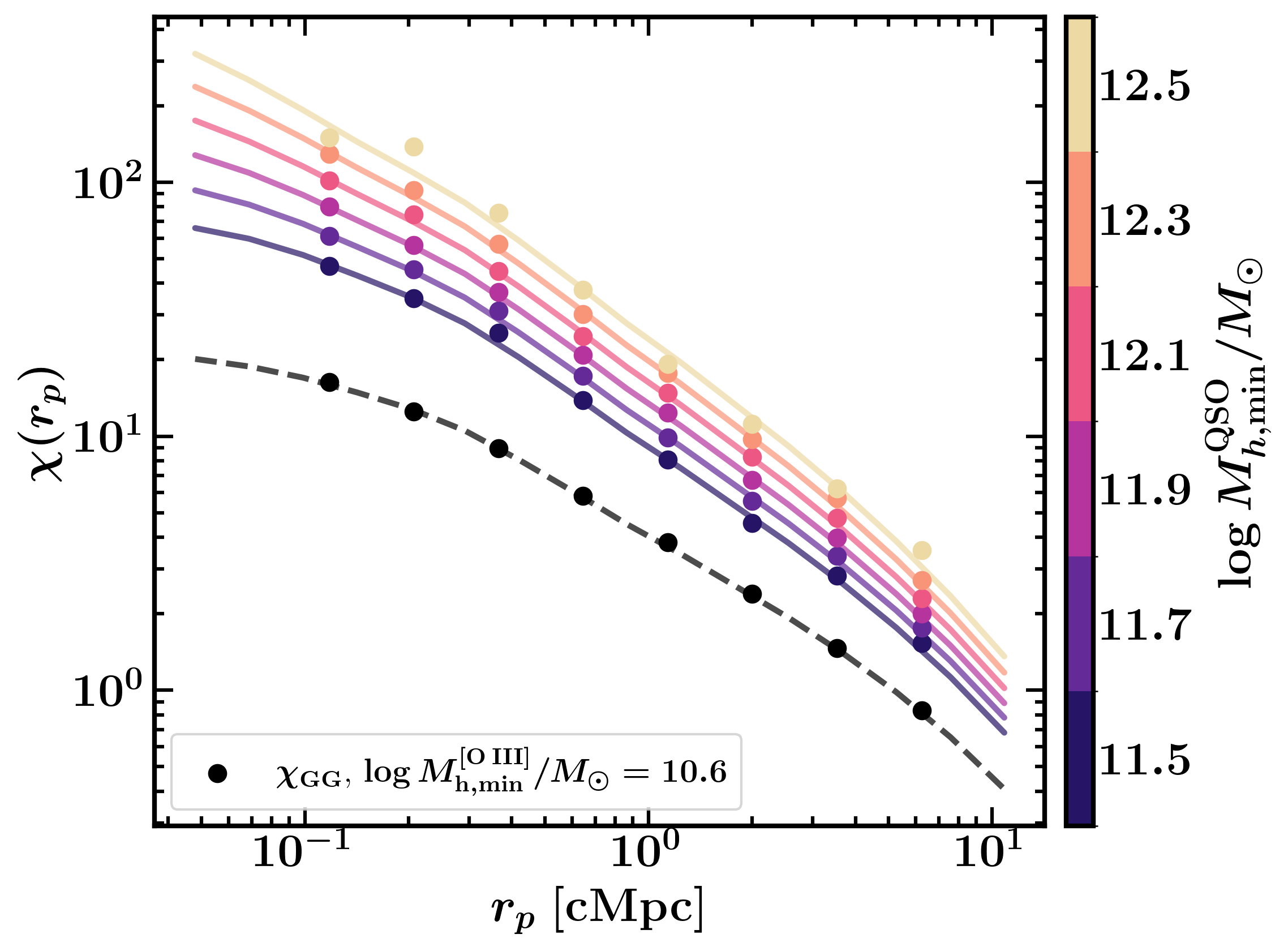}
        \label{fig:qg_corr}
    \end{subfigure}
\caption{\textbf{\textbf{\textit{Left:}}} Modeled \OIII-emitter auto-correlation function. The solid curves represent the results from the full simulated catalog, while scatter points show the mean of 1,000 ASPIRE-like realizations. The colors correspond to step-function halo occupation distributions with $\log\mming/\msun = 10.5$--$11.1$. \textbf{\textit{Right:}} QSO--galaxy cross-correlation function with $\log\mminq/\msun = 11.5$--$12.5$ and a fixed $\log\mming/\msun = 10.7$. The dashed line shows the reference \OIII\ auto-correlation function $\chi_{\rm GG}$ at fixed $\log\mming/\msun = 10.7$ for comparison.}
\label{fig:combined_corr}
\end{figure*}

To construct the random catalog, we employ a Monte Carlo approach. For each ASPIRE pointing, $5 \times 10^{6}$ mock sources are placed uniformly in RA and Dec within the F356W imaging footprint. Each source is assigned a redshift and \OIII\ luminosity drawn from the \OIII\ luminosity function measured by the EIGER program \citep{Matthee2023}, which covers the same redshift range ($5.3 < z < 7.0$) and WFSS mode as ASPIRE. Sources are then filtered through the coverage map (\S\ref{sec:coverage}) to verify geometric observability and through the sensitivity map (\S\ref{sec:sensitivity}) to verify detectability above the local flux limit. The resulting effective completeness, defined as the fraction of injected sources that survive both selection cuts, is $\langle S \rangle \approx 0.54$. We note that this completeness depends on both the flux limit of the survey and the area of the Monte Carlo injection region relative to the redshift-dependent WFSS spectral coverage. The WFSS coverage varies with redshift because the dispersed location of the \OIII\ lines 
shifts across the detector, changing the effective area observed by the two NIRCam modules. Since the injection region encompasses the
full rectangular F356W direct imaging footprint, which is larger than the spectral coverage at any given redshift, a fraction of injected sources fall outside the observable area. The effective completeness restricted to the WFSS-covered area, which is the relevant quantity for clustering measurements, is substantially higher; we refer the reader to \citet{Huang2026} for the explicit definition and value. 
The Landy--Szalay estimator normalizes the pair counts using $N_{R_{\rm G}}$, the expected number of \OIII-emitters in a blank field after applying the ASPIRE selection function, which is obtained by downscaling the oversampled random catalog by the effective completeness $\langle S \rangle$ (see \citealt{Huang2026} for details).

\section{Covariance matrix based on mock realizations}
\label{sec:cov}
\subsection{Simulation Setup}
Previous high-redshift clustering studies commonly estimate uncertainties using Poisson errors on the pair counts. However, because galaxies contribute to multiple pair-count bins and because large-scale density fluctuations (cosmic variance) coherently affect all bins, the true covariance matrix has significant off-diagonal structure that Poisson errors neglect entirely. The small number of independent ASPIRE fields makes the empirical or data-driven 
error estimators such as jackknife unreliable \citep{Norberg2009}. We therefore construct mock covariance matrices from the FLAMINGO-10k (F10k) simulation, a dark-matter-only (DMO) run from the FLAMINGO suite of cosmological simulations \citep{Schaye2023, Kugel2023} in a $(2.8\,\mathrm{cGpc})^3$ volume (Schaller et al., in prep.; see also \citealt{Pizzati2024b}). The simulation adopts the ``$3\times2$pt + all'' cosmology from \citet{Abbott2022}: $\Omega_{\rm m} = 0.306$, $\Omega_{\rm b} = 0.0486$, $\sigma_8 = 0.807$, $H_0 = 68.1\,{\rm km\,s^{-1}\,Mpc^{-1}}$, $n_s = 0.967$, with a summed neutrino mass of 0.06~eV. The simulation contains $10080^3$ CDM particles and $5600^3$ neutrino particles, with a CDM particle mass of $M_{\rm dm} = 8.40 \times 10^8\,\msun$.

Halos are identified using the HBT-HERONS (Hierarchical Bound-Tracing) halo finder \citep{Han2012, Han2018, ForouharMoreno2025}, which tracks the formation and evolution of subhalos by following their bound particles across cosmic time. We define the subhalo mass as the peak bound mass ($M_{\rm peak}$), i.e.\ the maximum mass a subhalo has attained over its history, and adopt the most bound particle as the subhalo position. The HBT-HERONS catalog includes both central and satellite subhalos. We use the simulation snapshot at $z_{\rm snap} = 6.14$, the closest match to the median redshift of \OIII-emitters observed in ASPIRE ($\langle z_{\rm \OIII} \rangle = 6.13$). 
We only select subhalos with more than 40 dark matter particles, corresponding to a minimum resolved mass of $\log(M^{\rm res}_{\rm min}/\msun) = 10.5$ \citep{Pizzati2024b}. See \citet{Pizzati2024b} for a detailed description of the halo catalog.

We adopt a step-function halo occupation distribution \citep{Zheng2005} in which both quasars and \OIII-emitters populate all subhalos (centrals and satellites) above their respective minimum mass thresholds, $\mminq$ and $\mming$. To match the observed abundance, we down-sample the galaxy catalog to reproduce the number density measured in \citet{Matthee2023}. Each selected halo is assigned an \OIII\ luminosity drawn from the EIGER luminosity function at $z \sim 5.3$--7 \citep{Matthee2023}, which is necessary because the mock sources are subsequently passed through the sensitivity map (\S\ref{sec:sensitivity}) to determine detectability. The full mock construction procedure is described in \S\ref{sec:mock_construction}.


\begin{figure*}
\centering
    \begin{minipage}{0.24\textwidth}
        \centering
        \textbf{$\log (\mminq/\msun) = 11.0$} \\
        \includegraphics[width=\textwidth]{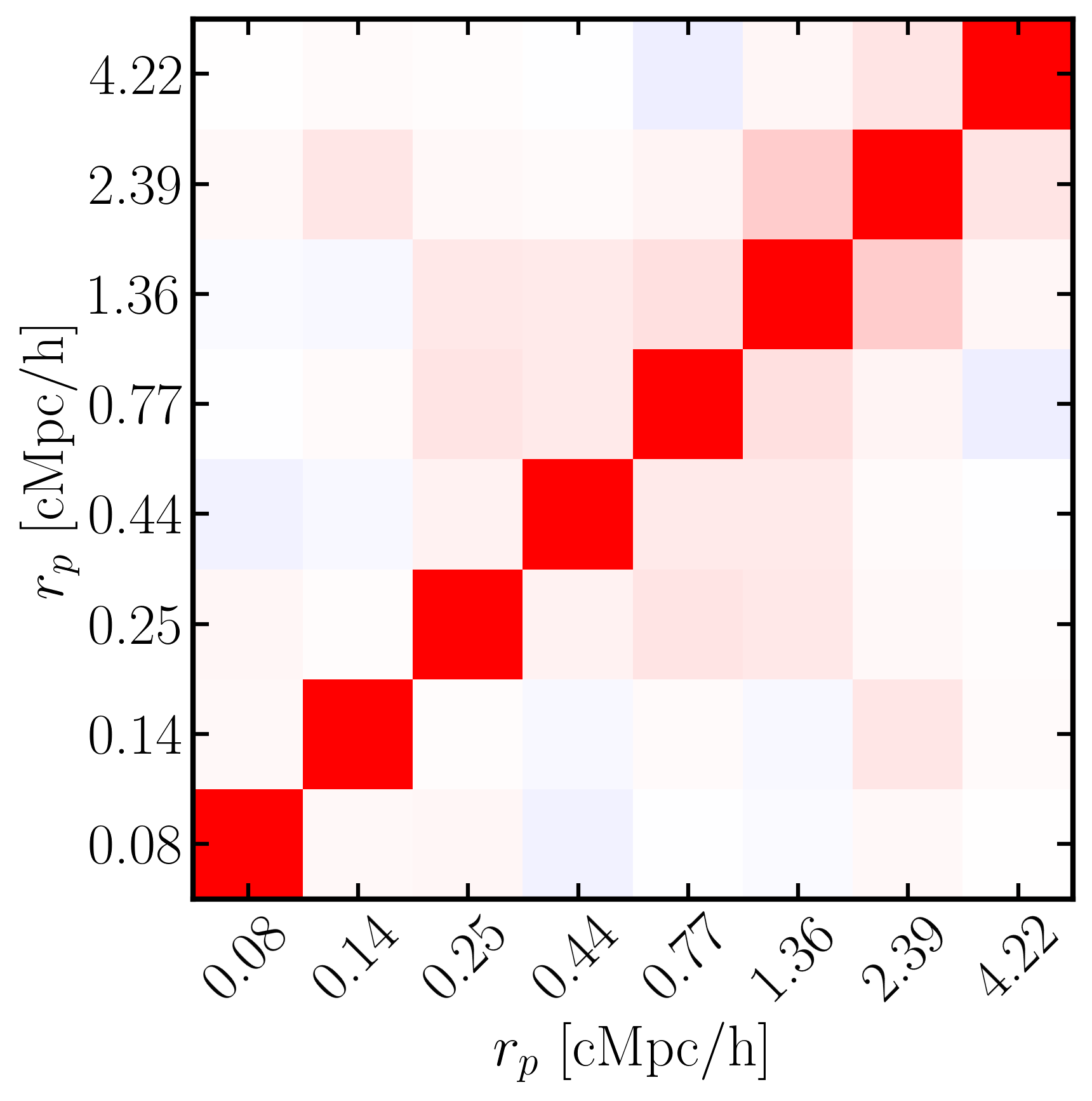}
    \end{minipage}%
    \begin{minipage}{0.24\textwidth}
        \centering
        \textbf{$\log (\mminq/\msun) = 11.5$} \\
        \includegraphics[width=\textwidth]{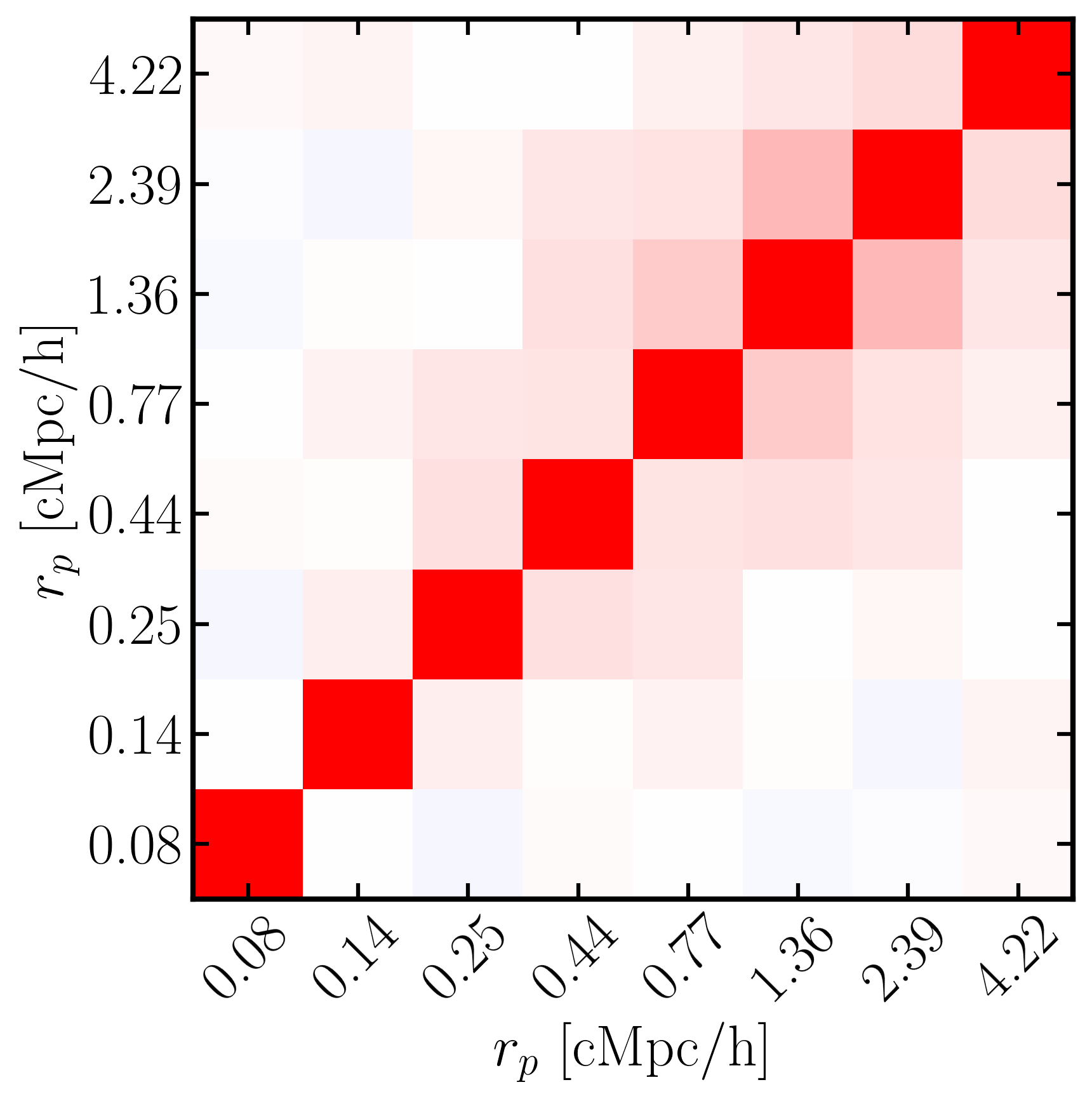}
    \end{minipage}%
    \begin{minipage}{0.24\textwidth}
        \centering
        \textbf{$\log (\mminq/\msun) = 12.0$} \\
        \includegraphics[width=\textwidth]{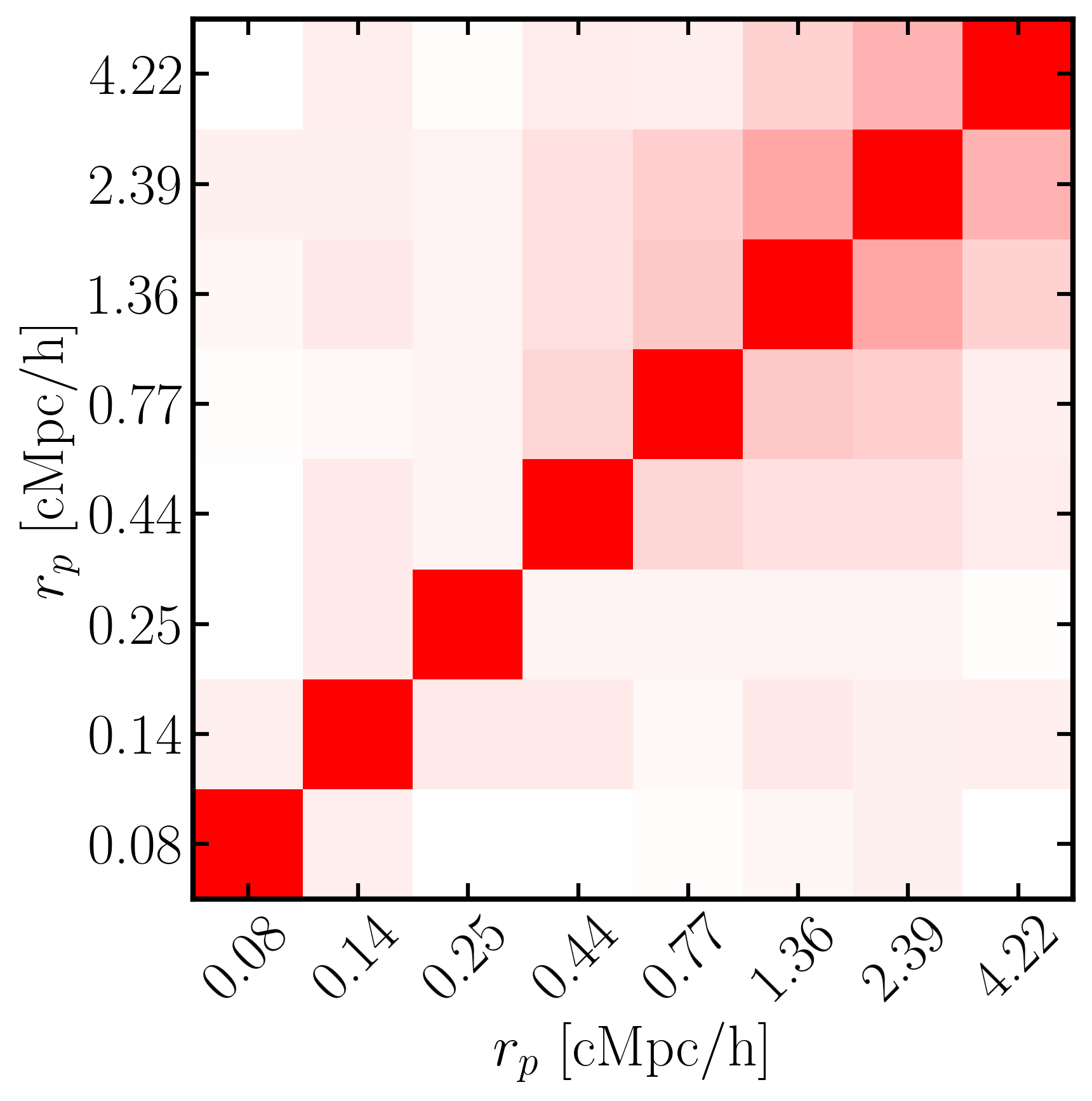}
    \end{minipage}%
    \begin{minipage}{0.24\textwidth}
        \centering
        \textbf{$\log (\mminq/\msun) = 12.5$} \\
        \includegraphics[width=\textwidth]{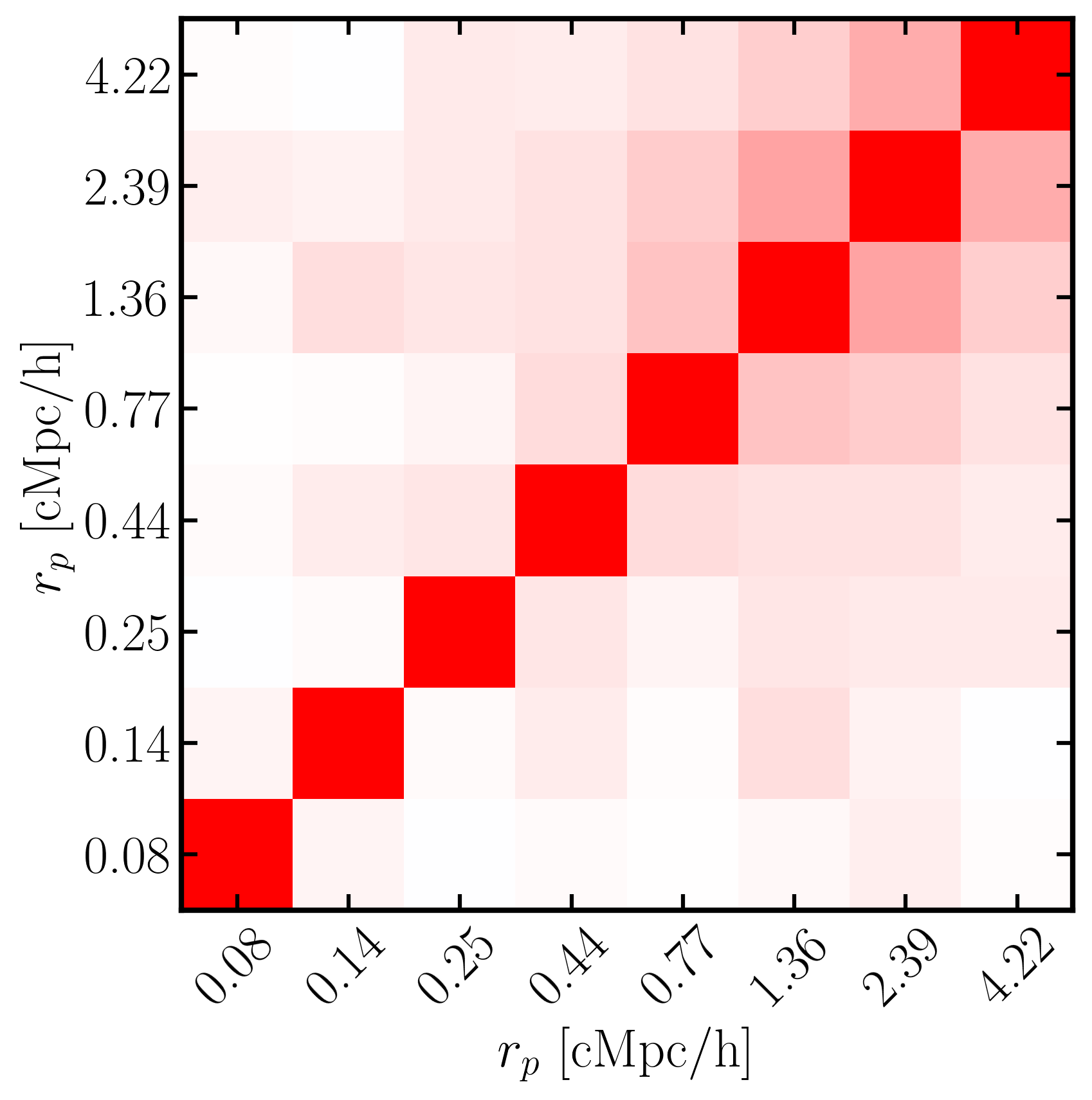}
    \end{minipage}
    \caption{Example correlation structure of the volume-averaged cross correlation function, $\chi_{\rm QG}$, computed using Eq.~\ref{eq:correlation_matrix} based on 1,000 realizations. From left to right, the minimum quasar host halo mass, $\mminq/\msun$, increases from 11.0 to 12.5, while the minimum galaxy host halo mass is fixed at $\log (\mming/\msun) = 10.6$ for better comparison. The off-diagonal correlation becomes slightly stronger as $\mminq/\msun$ increases, indicating stronger correlations for more massive quasar host halos.}
    \label{fig:cov_cross}
\end{figure*}

\begin{figure*}
\centering
    \begin{minipage}{0.24\textwidth}
        \centering
        \textbf{$\log (\mming/\msun) = 10.6$} \\
        \includegraphics[width=\textwidth]{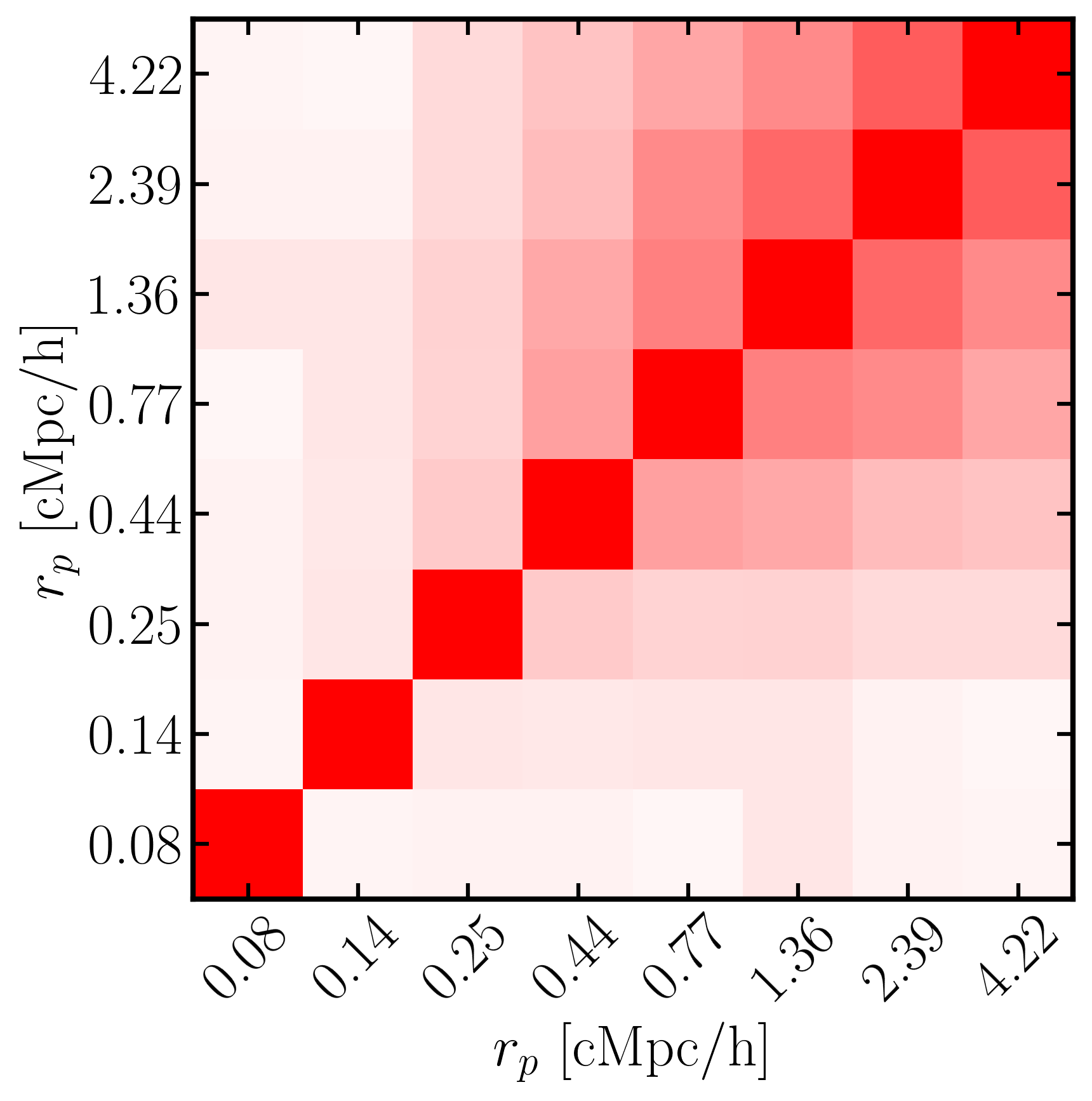}
    \end{minipage}%
    \begin{minipage}{0.24\textwidth}
        \centering
        \textbf{$\log (\mming/\msun) = 10.7$} \\
        \includegraphics[width=\textwidth]{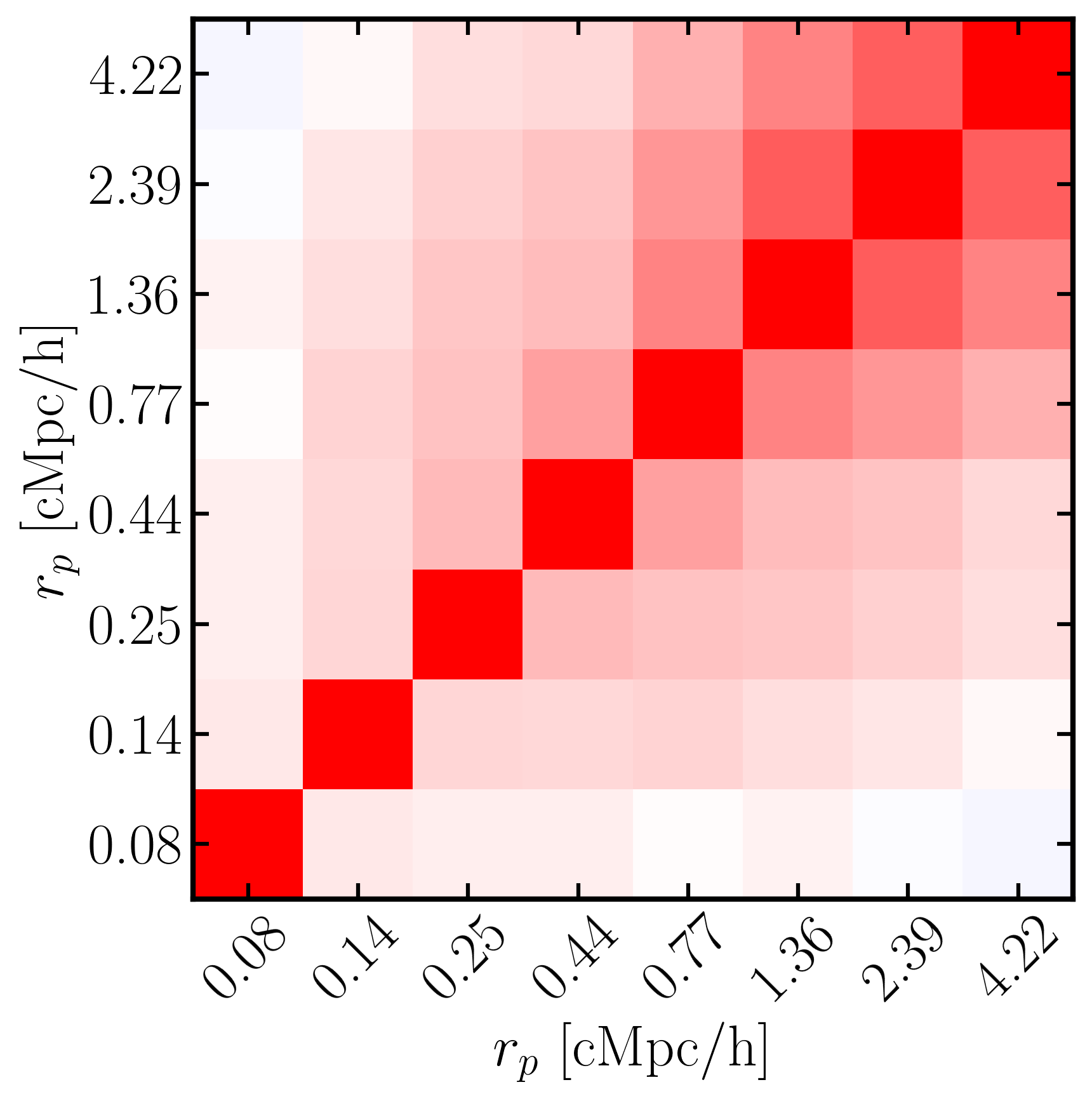}
    \end{minipage}%
    \begin{minipage}{0.24\textwidth}
        \centering
        \textbf{$\log (\mming/\msun) = 10.8$} \\
        \includegraphics[width=\textwidth]{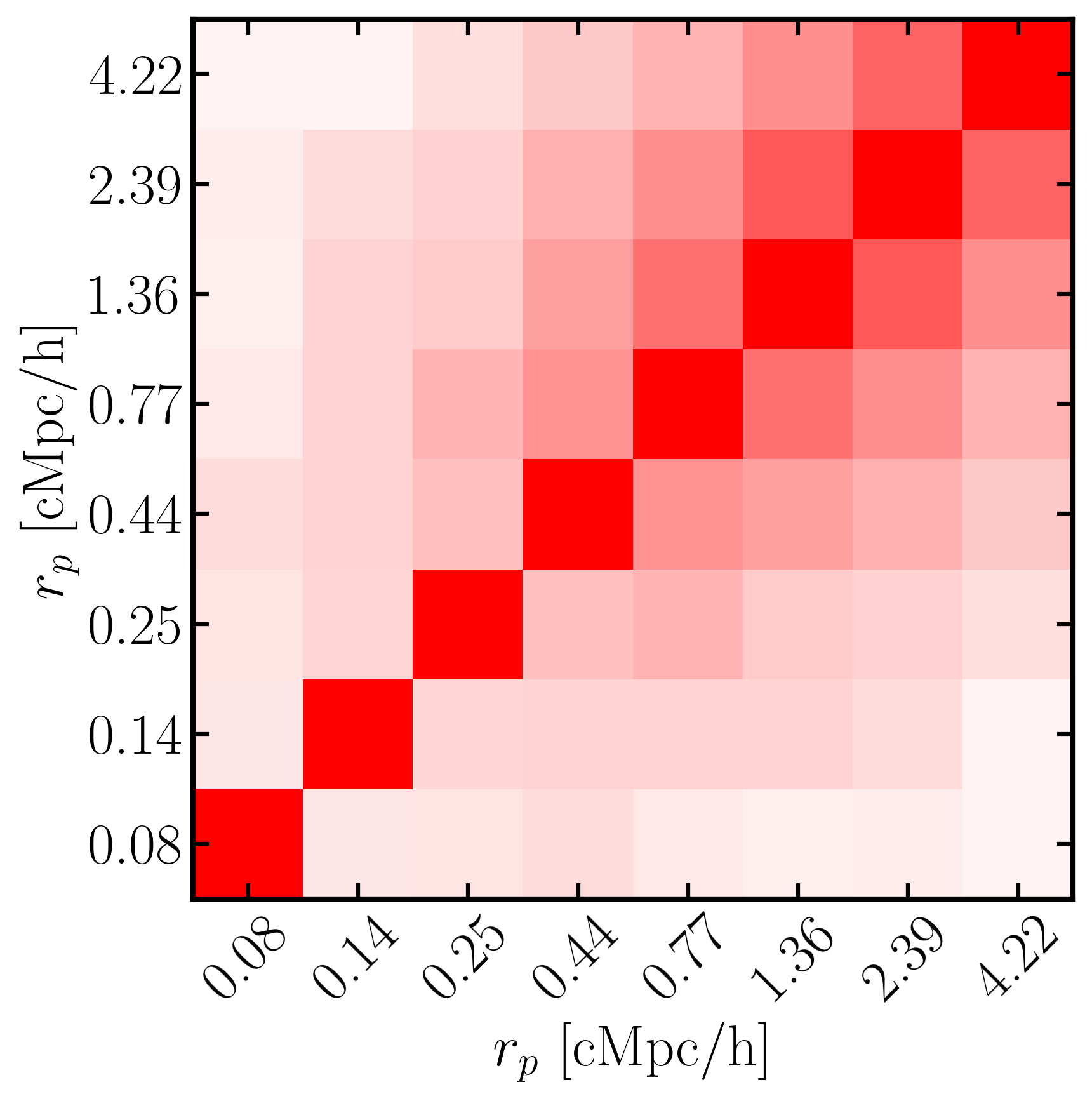}
    \end{minipage}%
    \begin{minipage}{0.24\textwidth}
        \centering
        \textbf{$\log (\mming/\msun) = 10.9$} \\
        \includegraphics[width=\textwidth]{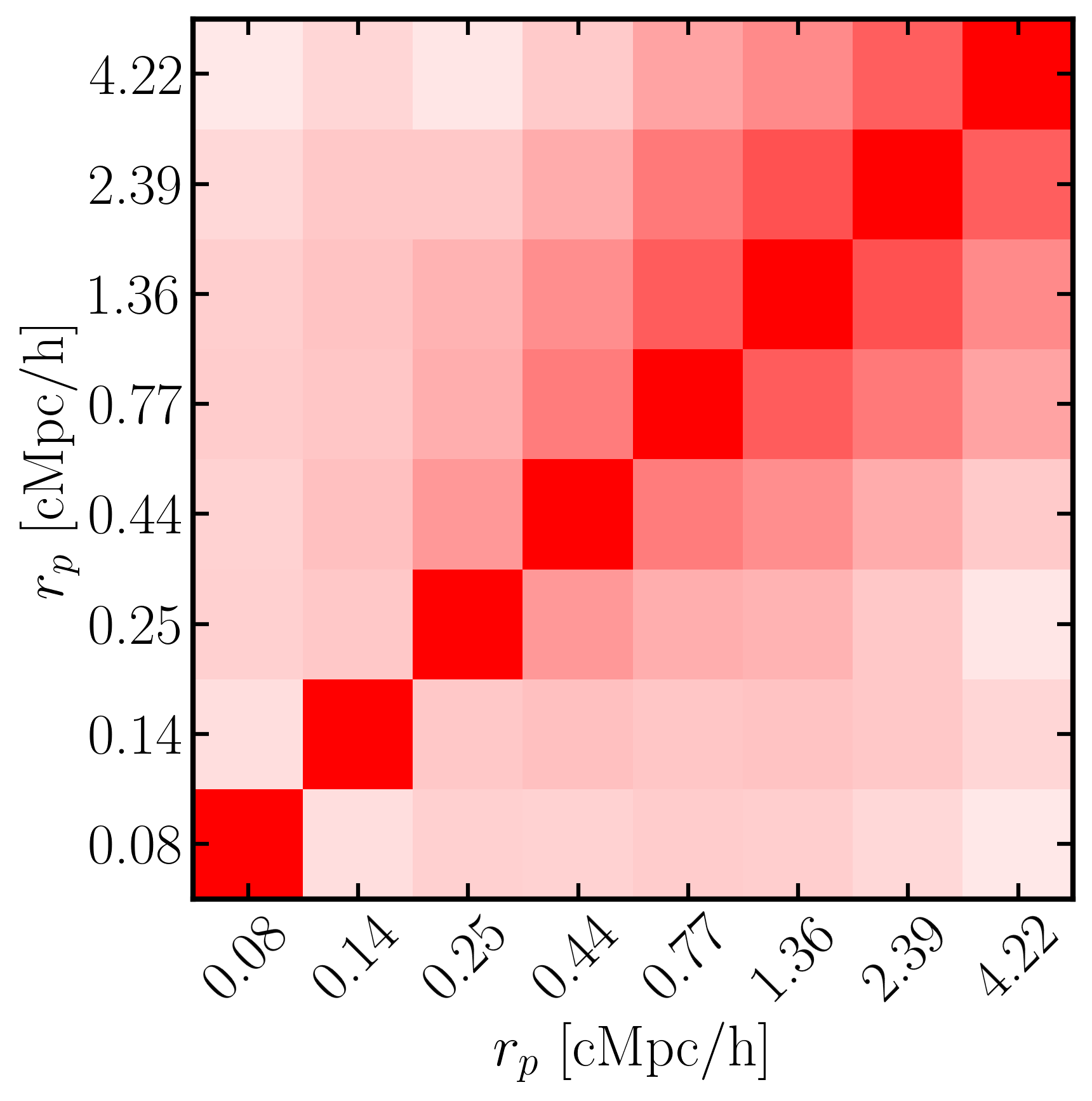}
    \end{minipage}

    \begin{minipage}{0.5\textwidth}
        \centering
        \includegraphics[width=\textwidth]{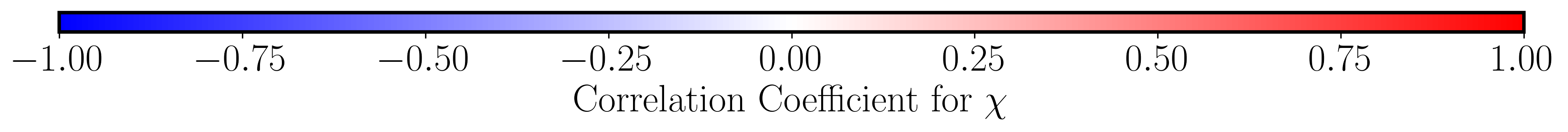}
    \end{minipage}
    
    \caption{Correlation structure of the volume-averaged auto-correlation function, $\chi_{\rm GG}$, computed using Eq.~\ref{eq:correlation_matrix} based on 1,000 realizations. From left to right, the minimum galaxy host halo mass, $\mming/\msun$ increases from 10.6 to 10.9. The off-diagonal correlation becomes slightly stronger as $\mming/\msun$ increases.}
    
    \label{fig:cov_auto}
    
\end{figure*}

\subsection{Construction of Mock \texorpdfstring{\OIII}{[O III]}-emitter Catalog}
\label{sec:mock_construction}
Each ASPIRE pointing covers approximately $2.2'\times 5.1'$ on the sky, corresponding to a transverse comoving size of $\approx 3.8 \times 8.8\,\mpch$ at $z \simeq 6.6$, with a line-of-sight depth of $\approx 485\,\mpch$ over $z = 5.3$--$7.0$. All three dimensions are much smaller than the $\approx 1900\,\mpch$ FLAMINGO-10k box side, so the simulation hosts many independent ASPIRE-like realizations from non-overlapping regions.

For each pair of minimum peak subhalo masses $(\mminq, \mming)$, we construct mock ASPIRE-like observations as follows:
\begin{enumerate}
    \item Select quasars and \OIII-emitters from the simulation using a step-function HOD: all subhalos (both centrals and satellites) with $M_{\rm peak} > \mminq$ are assigned as quasar hosts, and all subhalos with $M_{\rm peak} > \mming$ as galaxy hosts. The positions of both quasars and galaxies are taken to be the most bound particle of their host subhalo. We randomly down-sample the galaxy catalog to match the observed \OIII-emitter number density from \citet{Matthee2023}; the down-sampling factor corresponds to the \OIII-emitter duty cycle, $f^{\rm \OIII}_{\rm duty} = n_{\rm \OIII}/n_{M_{\rm h} > \mming}$. This assumes a common mass threshold for centrals and satellites; the impact of relaxing this assumption is explored in \S\ref{sec:split_hod}. We adopt $\log(\mminq/\msun) = 11.0$--$12.5$ and $\log(\mming/\msun) = 10.5$--$11.0$, bracketing the values inferred from the EIGER survey \citep{Pizzati2024b}.

    \item For each realization, randomly select 25 quasars from the catalog. Using the periodic boundary conditions of the simulation box, re-center the galaxy catalog around each selected quasar to avoid edge effects. Assign observed sky coordinates and redshifts by mapping the simulation-frame positions relative to each quasar onto the true ASPIRE quasar locations. Galaxy redshifts include the line-of-sight peculiar velocity from the simulation, encoding redshift-space distortions \citep[see][for the explicit redshift-assignment procedure]{Huang2026}. Each galaxy is assigned an \OIII\ luminosity drawn randomly from the EIGER luminosity function at $z \sim 5.3$--$7$ \citep{Matthee2023}, independently of its host halo mass and assuming no redshift evolution over this range.

    \item Apply the redshift-dependent coverage map (\S\ref{sec:coverage}) and sensitivity map (\S\ref{sec:sensitivity}) field by field to determine which mock \OIII-emitters would be detected in the ASPIRE survey.

    \item Compute the \OIII-emitter auto-correlation function $\chi_{\rm GG}$ (Eq.~\ref{eq:auto}) and the quasar--\OIII-emitter cross-correlation function $\chi_{\rm QG}$ (Eq.~\ref{eq:cross}), based on the total pair counts across all 25 mock ASPIRE fields.

    \item Repeat steps (ii)--(iv) for $N_{\rm real} = 1000$ realizations and compute the covariance matrix using Eq.~\ref{eq:covariance}.
\end{enumerate}

Figure~\ref{fig:combined_corr} compares the mean of the 1000 mock correlation functions (scatter points) with the full-box halo model predictions (solid curves) for both $\chi_{\rm GG}$ and $\chi_{\rm QG}$ across a range of minimum halo masses, confirming that the ASPIRE-like mocks faithfully recover the input clustering signal. From these $N_{\rm real} = 1000$ realizations, we compute the covariance matrix elements as:

\begin{equation}
\mathcal{C}_{ij}(\chi_a)=
\frac{1}{N_{\rm real}}
\sum_{k=1}^{N_{\rm real}}
\bigl(\chi_{a,i}^{k}-\langle \chi_{a,i}\rangle\bigr)
\bigl(\chi_{a,j}^{k}-\langle \chi_{a,j}\rangle\bigr),
\label{eq:covariance}
\end{equation}
where indices $i,j=1,\dots,N_{\rm bin}$ label radial bins, $k=1,\dots,N_{\rm real}$ labels realizations, $\chi_{a,i}^{k}$ is the $i$-th bin for the $k$-th mock realization of the volume-averaged correlation function, $a\in\{\rm GG, QG\}$, and $N_{\rm real}=1000$ is the total number of realizations. To visualize the bin-to-bin correlations, we define the correlation matrix:
\begin{equation}
\rho_{ij}(\chi_a) = \frac{\mathcal{C}_{ij}(\chi_a)}{\sqrt{\mathcal{C}_{ii}(\chi_a)\,\mathcal{C}_{jj}(\chi_a)}}.
\label{eq:correlation_matrix}
\end{equation}

Figs.~\ref{fig:cov_cross} and \ref{fig:cov_auto} show the resulting correlation matrices (Eq.~\ref{eq:correlation_matrix}) for the cross-correlation and auto-correlation, respectively. The cross-correlation correlation matrix (Fig.~\ref{fig:cov_cross}) shows weak off-diagonal terms, since $\chi_{\rm QG}$ is measured around a single quasar per field and each galaxy contributes to only one separation bin relative to the quasar. The off-diagonal elements increase with $\mminq$: for more massive, more strongly biased quasar host halos, large-scale density fluctuations contribute a larger fraction of the total variance relative to shot noise, increasing the correlation between radial bins \citep[see, e.g.,][]{Bernstein1994}. The auto-correlation correlation matrix (Fig.~\ref{fig:cov_auto}) exhibits stronger off-diagonal correlations, which increase with $\mming$. This arises from two effects: (1) the same galaxy contributes to pair counts at multiple separations, so uncertainties in different bins are inherently coupled; and (2) large-scale density fluctuations coherently boost or suppress galaxy counts across all bins simultaneously.


\section{Statistical Inference with Covariance Matrices}
\label{sec:mcmc}
\subsection{Likelihood functions}
\label{sec:likelihood}
By construction, $\chi_{\rm QG}$ is computed using only \OIII-emitters within $\pm 7\,\mpch$ of the quasar along the line of sight (i.e., the same $\pi_{\max}$ window used in the cross-correlation estimator), while $\chi_{\rm GG}$ uses only galaxies outside that region. The two estimators therefore use disjoint pair sets, and we have verified directly from the 1000 mock realizations that the cross-covariance between $\chi_{\rm GG}$ and $\chi_{\rm QG}$ is negligible. We therefore treat the two correlation functions as statistically independent. The total likelihood function is:
\begin{equation}
\mathcal{L}_{\text{tot}}(X \mid \Theta) = \mathcal{L}_{\text{GG}} \cdot \mathcal{L}_{\text{QG}},
\label{eq:total_likelihood}
\end{equation}
where $X$ is a vector of observed correlation functions, $X = (\chi_{\text{GG}}, \chi_{\text{QG}})$ and $\Theta$ is the modeled parameters for the minimum masses of subhalos hosting quasars and galaxies,
$\Theta = (\mminq, \mming)$.
The individual likelihoods $\mathcal{L}_{\text{GG}}$ and $\mathcal{L}_{\text{QG}}$ are:

\begin{equation}
\mathcal{L}_{\rm GG} = \mathcal{N}\left(\chi_{\rm GG}^{\rm obs} \middle| \chi_{\rm GG}^{\rm model}(\mming), \boldsymbol{C}_{\rm GG}(\mming)\right)
\label{eq:auto_likelihood}
\end{equation}
and
\begin{equation}
\mathcal{L}_{\text{QG}} = \mathcal{N}\left(\chi_{\rm QG}^{\rm obs} \middle| \chi_{\rm QG}^{\rm model}(\mminq, \mming), \boldsymbol{C}_{\rm QG}(\mminq, \mming)\right).
\end{equation}

Here, $\mathcal{N}(\chi^{\rm obs} | \chi^{\rm model}, \boldsymbol{C})$ represents a Gaussian distribution with mean $\chi^{\rm model}$ and covariance $\boldsymbol{C}$, evaluated at the observed data $\chi^{\rm obs}$. The model prediction $\chi^{\rm model}$ is taken from the FLAMINGO-10k clustering fit of \citet{Pizzati2024b}, which is consistent with the mean of our 1000 ASPIRE-like mock realizations for $\log(M_h/\msun) \geq 10.5$ (Fig.~\ref{fig:combined_corr}). We use this fit rather than the mock mean directly because the inference grid extends below the FLAMINGO-10k resolution limit ($\log(M_h/\msun) = 10.5$); the fit can be extrapolated to lower masses, although the resulting prediction may not be exact in that regime.

For comparison, we also evaluate the likelihood using Poisson pair-count uncertainties:                                                          
\begin{equation}                                                                                               \label{eq:poisson_likelihood}                      
\log \mathcal{L}_{a}^{\rm Pois} = -\frac{1}{2} \sum_{i=1}^{N_{\rm bin}} \frac{(\chi_{a,i}^{\rm obs} - \chi_{a,i}^{\rm model})^2}{\sigma_{a,i}^2},                                                                                                              
\end{equation}
where $a \in [\rm GG, QG]$ and $\sigma_{a,i}$ is the Poisson error in bin $i$, obtained by propagating the $1\sigma$ Poisson confidence interval on the raw data--data pair counts through the correlation function estimator. For the \OIII-emitter auto-correlation, we propagate the Poisson interval on $D_{\rm G}D_{\rm G}$ through the Landy--Szalay estimator (Eq.~\ref{eq:auto}); for the cross-correlation, we propagate the interval on $D_{\rm Q}D_{\rm G}$ through the Davis--Peebles estimator (Eq.~\ref{eq:cross}). In both cases, the data--random and random--random pair counts are treated as noiseless, since the random catalog is oversampled to $5 \times 10^6$ sources per field (see \citealt{Huang2026} for details on the pair count normalization). When fitting a single mock realization, we use the Poisson error computed from the pair counts of that realization, as one would with real observational data. When the ``observed'' data vector is instead set to the mean of all realizations, we correspondingly use the mean Poisson error across all 1000 realizations (see detailed discussion in \S\ref{sec:cov_vs_poisson}). This allows us to directly quantify the impact of adopting the full covariance matrix on the inferred parameters.

\subsection{Minimum Halo Masses}

\label{sec:mmin}
We now apply the framework described above to infer the minimum halo masses $(\mminq, \mming)$, using the simulation-based clustering models and mass-dependent covariance matrices constructed in \S\ref{sec:cov}. Unlike the power-law correlation length analysis presented in Appendix~\ref{appendix:corr_length}, where the covariance must be fixed at a fiducial mass, here both the model correlation function and the covariance matrix depend on the fitted parameters $(\mminq, \mming)$.

\subsubsection{Mass Dependent Covariance Matrices}
Since the correlation functions $\chi_{\rm QG}$ and $\chi_{\rm GG}$ depend on the minimum halo mass threshold, so does their variance in each radial bin. As shown in Fig.~\ref{fig:cov_cross} and \ref{fig:cov_auto}, the off-diagonal covariance structure also varies with the minimum halo mass for both $\chi_{\rm QG}$ and $\chi_{\rm GG}$. To account for this mass dependence, we compute the covariance matrix on a grid of minimum halo masses: for $\chi_{\rm GG}$, six values of $\log\mming/\msun \in [10.5, 11.0]$ in 0.1-dex steps; for $\chi_{\rm QG}$, a 2D grid of 30 mass pairs spanning $\log\mming/\msun \in [10.5, 10.9]$ in 0.1-dex steps and $\log\mminq/\msun \in [11.5, 12.5]$ in 0.2-dex steps.

\begin{figure*}
\centering
    \includegraphics[width=0.48\textwidth]{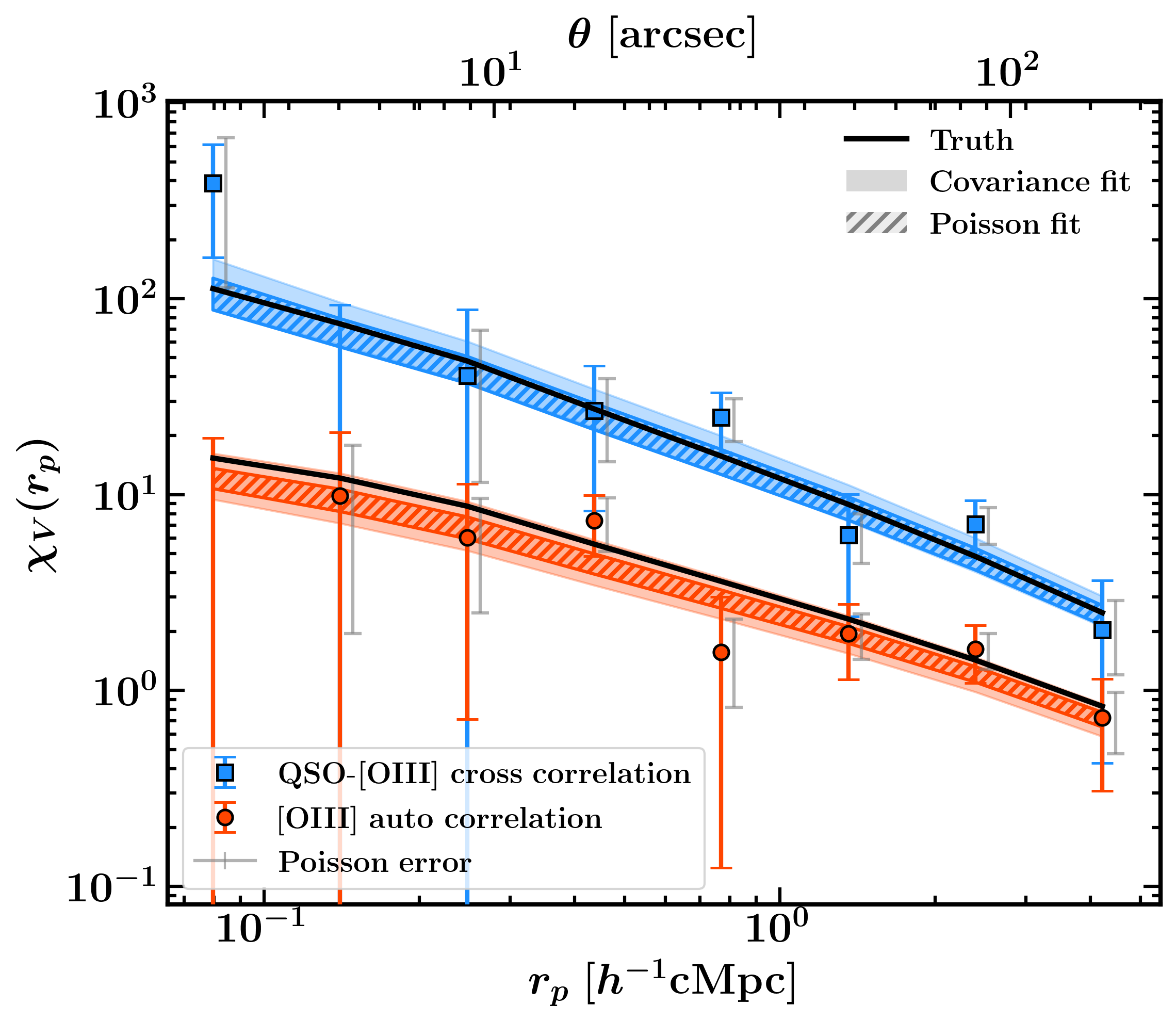}
    \hfill
    \includegraphics[width=0.48\textwidth]{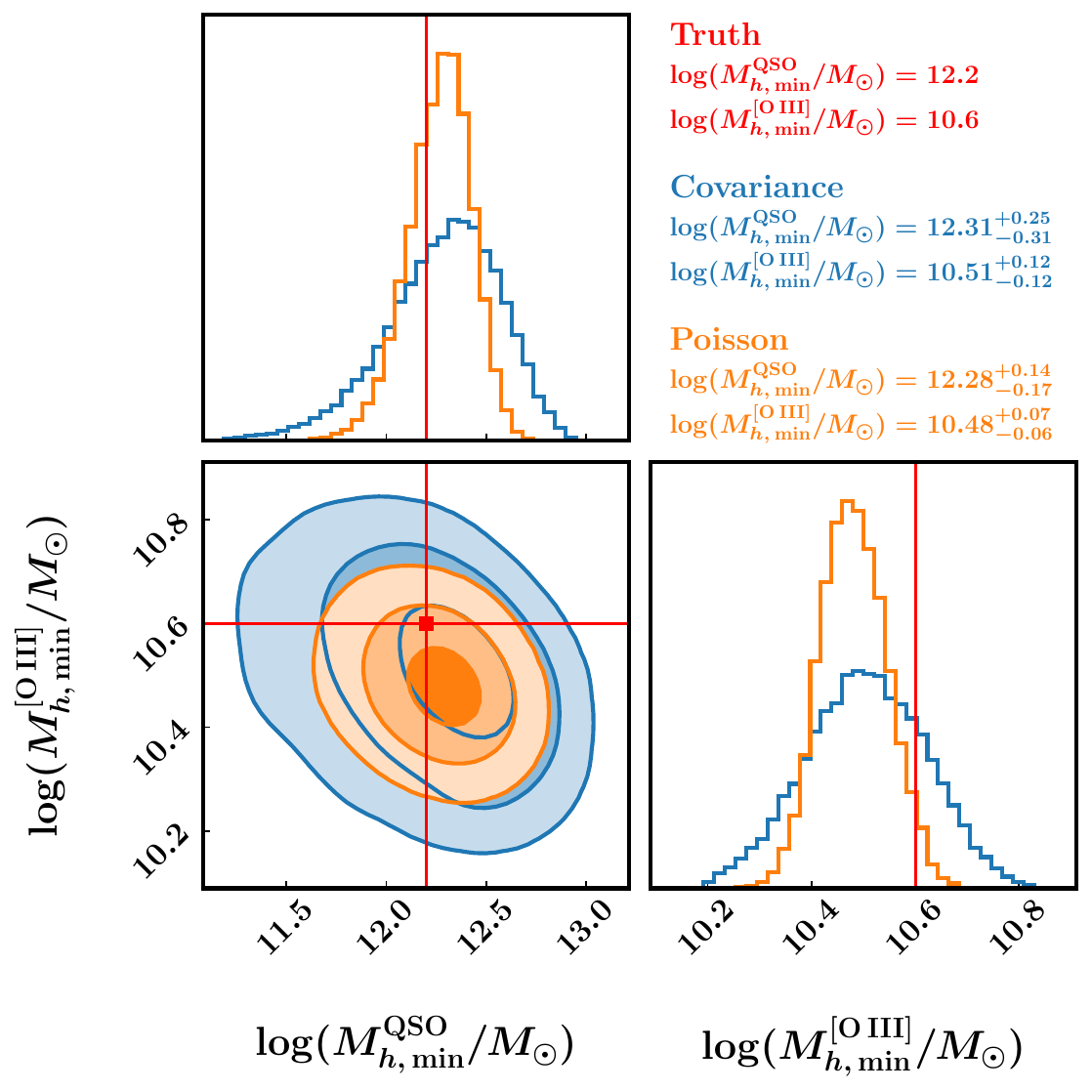}
    \caption{\textbf{\textit{Left:}} Correlation function fit for a single mock realization with input masses $\log(\mminq/\msun) = 12.2$ and $\log(\mming/\msun) = 10.6$. Blue squares show the quasar--\oiii\ cross-correlation and red circles show the \oiii\ auto-correlation. Colored error bars show the $1\sigma$ uncertainty from the diagonal of the full covariance matrix; grey offset error bars show the Poisson errors from pair counts for comparison. Black solid curves show the true model at the input masses. Solid shaded bands show the 16th--84th percentile model range from 200 posterior draws using the full covariance likelihood; hatched bands show the same for the Poisson likelihood. The top axis shows the corresponding angular scale at $z = 6.1$.
    \textbf{\textit{Right:}} Joint posterior distribution for $\mminq$ and $\mming$. Blue filled contours show the 1, 2, and $3\sigma$ posterior using the full mock covariance matrix; orange filled contours show the posterior using diagonal Poisson errors. Red crosshairs mark the true input values. The covariance-based posterior is $\approx 1.8\times$ broader in both parameters; the Poisson posterior excludes the true $\mming$ at $1\sigma$, illustrating the overconfidence of Poisson error estimates.
    }
    \label{fig:mock_inference}
\end{figure*}


\subsubsection{Example fit to a single mock realization}

To further illustrate the impact of the covariance matrix on parameter inference, we perform Bayesian inference and sample the posterior with MCMC
on individual mock realizations. We select one realization from our mock catalog generated with the fiducial input masses $\log(\mminq/\msun) = 12.2$ and $\log(\mming/\msun) = 10.6$, and treat its correlation functions $\chi_{\rm QG}(r_p)$ and $\chi_{\rm GG}(r_p)$ as ``observed'' data. 
We then run two independent MCMC inferences using the affine-invariant ensemble sampler \texttt{emcee} \citep{ForemanMackey2013}: one using the full mock covariance matrix in the likelihood (Equation~\ref{eq:total_likelihood}), and one replacing it with Poisson errors derived from the pair counts of that realization. Each chain uses 36 walkers run for 5000 steps, with the first 1000 steps discarded as burn-in.

The left panel of Fig.~\ref{fig:mock_inference} shows the resulting fits. The data points (blue squares for the quasar--\oiii\ cross-correlation; red circles for the \oiii\ auto-correlation) are plotted with colored error bars from the diagonal of the full covariance matrix, which represent the realistic $1\sigma$ uncertainties including cosmic variance. Grey offset error bars show the Poisson errors derived from pair counts for comparison, which are visibly smaller. The black solid curves show the true halo model evaluated at the input masses. The solid shaded bands show the 16th--84th percentile model range from 200 random draws of the covariance-based posterior, while the hatched bands show the same for the Poisson-based posterior.

The Poisson fit bands (hatched) are visibly narrower than the covariance fit bands (solid shaded), directly demonstrating that the Poisson likelihood produces artificially tight constraints. The two fits also yield slightly different median parameters: the Poisson fit gives $\log(\mminq/\msun) = 12.28^{+0.14}_{-0.17}$ versus $12.31^{+0.25}_{-0.31}$ for the covariance fit, and $\log(\mming/\msun) = 10.48^{+0.07}_{-0.06}$ versus $10.51^{+0.12}_{-0.12}$. This shift arises because the two likelihoods weight the radial bins differently. 

The right panel of Fig.~\ref{fig:mock_inference} shows the corresponding MCMC posterior distributions. We note that the posterior contours extend below the FLAMINGO-10k resolution limit of $\log\mming/\msun = 10.5$; in this regime $\chi^{\rm model}$ is the extrapolated model from \citet{Pizzati2024b} (see \S\ref{sec:likelihood}) and the inferred values should be interpreted with appropriate caution. The covariance-based posterior (blue contours) is approximately $1.8\times$ broader in both $\mminq$ and $\mming$ compared to the Poisson posterior (orange contours). For this realization, the true input values (red crosshairs) fall within the $1\sigma$ contour of the covariance-based posterior but are excluded by the Poisson posterior for $\mming$, illustrating the overconfidence of Poisson error estimates. We quantify this in \S\ref{sec:inference_test}: across 1000 mock realizations, the truth lies more than $1\sigma$ from the Poisson estimate in $\approx 80\%$ of cases, far above the $\approx 32\%$ expected for well-calibrated intervals. The single-realization example above demonstrates the effect of the full covariance at one fiducial mass pair. However, since the covariance matrix itself depends on the assumed halo masses, the degree to which Poisson errors underestimate the true uncertainty is expected to vary across the parameter space. In \S\ref{sec:cov_vs_poisson}, we extend this comparison across a grid of minimum halo masses to quantify this mass dependence. A complementary noise-free version of this single-realization fit, using the mean of all 1000 mock realizations as the observed data, is presented in Appendix~\ref{appendix:mean_mock}.

\begin{figure}
\centering
	\includegraphics[width=\columnwidth]{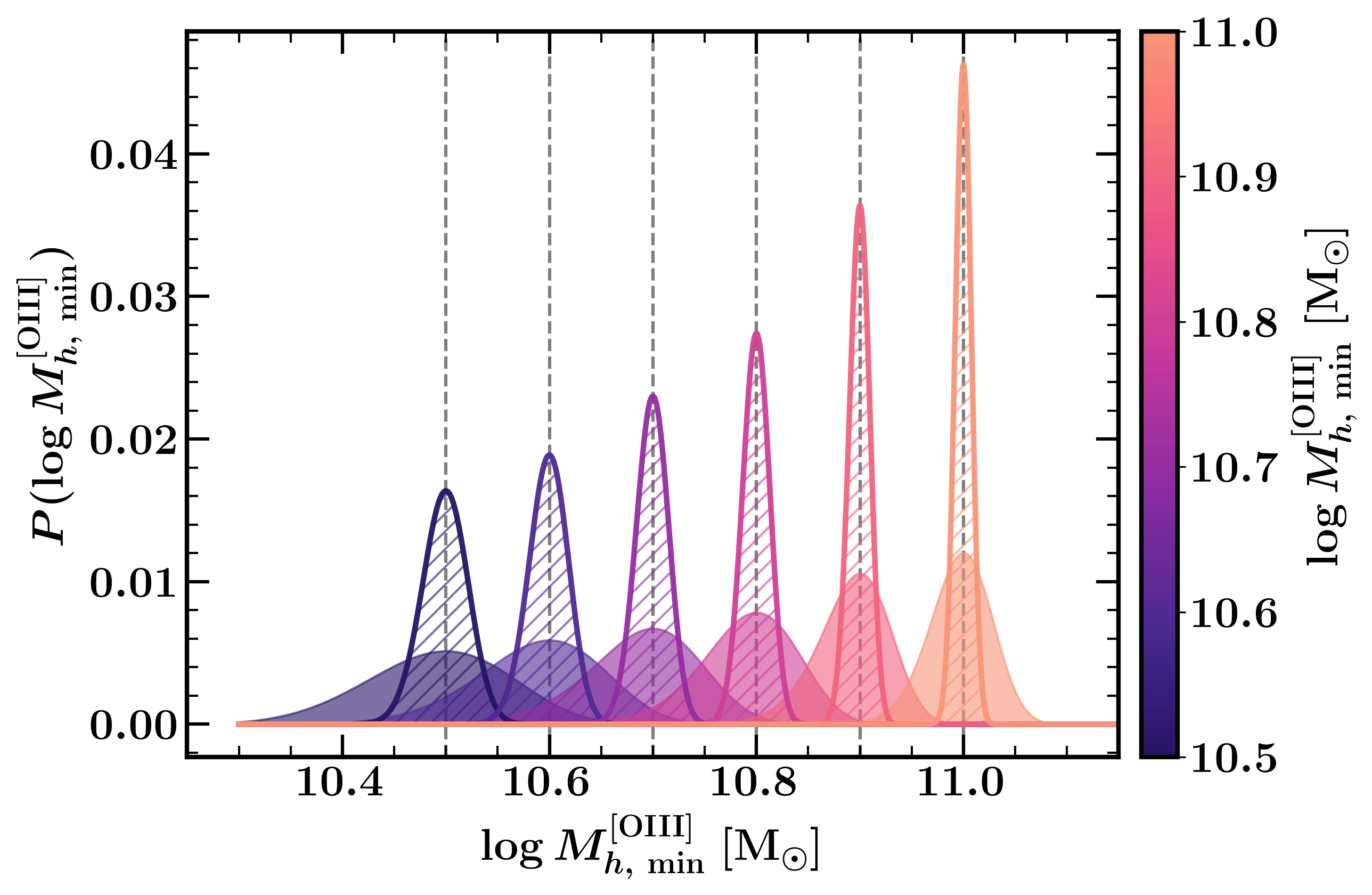}
    \caption{Normalized likelihood $P(\mming) \propto \mathcal{L}_{\rm GG}(\mming)$ evaluated on a grid of $\mming$ values, using only the GG auto-correlation function $\chi_{\rm GG}(r_p)$. Because $\chi_{\rm GG}$ depends only on $\mming$ and is independent of $\mminq$, no marginalization is required. For each of six input mass thresholds (dashed vertical lines), the ``observed'' data vector is set equal to the mean mock $\chi_{\rm GG}$ at the true $\mming$, so that both likelihoods peak at the correct input value by construction. The darker filled curve evaluates $\mathcal{L}_{\rm GG}$ using the full mock covariance $\boldsymbol{C}_{\rm GG}$, and the lighter filled curve uses diagonal Poisson errors. The difference between the two is therefore purely in the error budget: the covariance-based constraints are systematically broader, reflecting cosmic variance and off-diagonal bin-to-bin correlations that Poisson errors neglect.
    }
    \label{fig:likelihood_Mgal}
\end{figure}


\begin{figure}
\centering
	\includegraphics[width=\columnwidth]{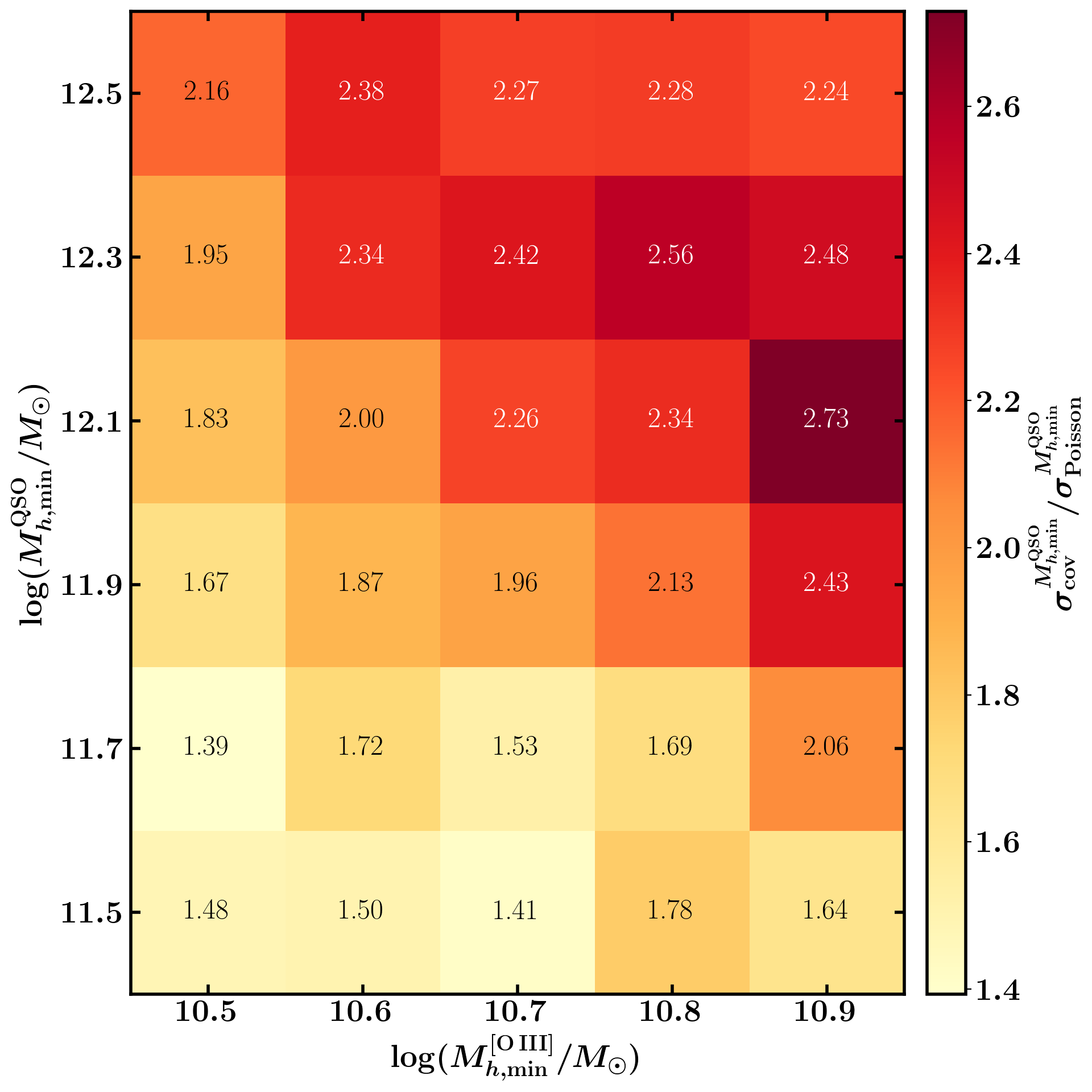}
    \caption{Ratio $\sigma^{\mminq}_{\rm cov}/\sigma^{\mminq}_{\rm Poisson}$ for the marginalized $\mminq$ constraint from the joint QG$+$GG fit, plotted across a grid of minimum halo masses for quasars ($\mminq$, $y$-axis) and galaxies ($\mming$, $x$-axis). The ratio increases with both masses, reaching $\approx$2.4--2.7 at the highest masses, reflecting the stronger cosmic variance for more biased tracers. At the highest quasar halo masses ($\log\mminq/\msun \geq 12.3$, top two rows), the ratio does not grow monotonically. This is because the mock covariance was generated only on $\log\mminq/\msun \in [11.5, 12.5]$, so inferences at higher input masses fall back to the $\mminq = 12.5$ covariance; and the model $\chi(r_p)$ grid extends only to $\log\mminq/\msun = 13.0$, truncating the posterior at this upper boundary. Because the covariance posterior is broader than the Poisson posterior, it is clipped more by this upper-boundary truncation, so $\sigma^{\mminq}_{\rm cov}$ shrinks faster than $\sigma^{\mminq}_{\rm Poisson}$. The ratio therefore stops growing — and in some cells decreases — at the top of the grid, instead of continuing to increase with $\mminq$ as it does at lower masses.
    }
    \label{fig:sigma_ratio_heatmap}
\end{figure}

\subsection{Impact of covariance on halo mass constraints}
\label{sec:cov_vs_poisson}

To quantify the importance of the full covariance matrix in constraining halo minimum masses, we compare parameter uncertainties derived using the full mock covariance $\boldsymbol{C}$ (Eq.~\ref{eq:total_likelihood}) against those obtained using the Poisson likelihood (Eq.~\ref{eq:poisson_likelihood}). To isolate the effect of the error model from noise in any particular realization, we set the ``observed'' data vector to the mean of 1000 mock realizations, i.e., the noiseless true model. This ensures that both likelihoods peak at the same input value by construction, so the difference between the two is purely in the width of the constraints. As described in \S\ref{sec:likelihood}, we use the mean Poisson error across all realizations as the representative Poisson uncertainty, consistent with using the noise-free mean as the observed data. Unlike the full covariance, this Poisson error captures only shot noise and ignores both cosmic variance and correlations between $r_p$ bins.



We generate the covariance matrices $\boldsymbol{C}_{\rm QG}$ and $\boldsymbol{C}_{\rm GG}$ from 1000 mock realizations of the ASPIRE survey. For the galaxy auto-correlation, we compute $\boldsymbol{C}_{\rm GG}$ independently at six values of the galaxy halo mass threshold, $\log(\mming/\msun) \in [10.5, 11.0]$ in steps of 0.1~dex. For the cross-correlation, mock covariances are computed 
on a 2d grid of 30 mass pairs spanning $\mming \in [10.5, 10.9]$ in steps of 0.1~dex and $\mminq \in [11.5, 12.5]$ in steps of 0.2~dex ($5 \times 6 = 30$ mass pairs).

For each $(\mming, \mminq)$ pair, 
we perform two parallel inferences: one using the full covariance matrix (Equation~\ref{eq:total_likelihood}) and one replacing $\boldsymbol{C}$ with the diagonal Poisson errors. From the resulting likelihood curves we compute the $1\sigma$ uncertainty (half the 16th--84th percentile interval of the normalized likelihood) and compute the ratio $\sigma^{\mminq}_{\rm cov} / \sigma^{\mminq}_{\rm Poisson}$.

Fig.~\ref{fig:likelihood_Mgal} shows the normalized likelihood $P(\mming) \propto \mathcal{L}_{\rm GG}(\mming)$ from the GG auto-correlation alone, evaluated on a grid of $\mming$ values for six input mass thresholds (dashed vertical lines). Because $\chi_{\rm GG}$ depends only on $\mming$ and is independent of $\mminq$, no marginalization over the quasar halo mass is required. As described at the beginning of \S\ref{sec:cov_vs_poisson}, the ``observed'' data vector is set to the mean mock $\chi_{\rm GG}$ at each true input mass, so both likelihoods peak at the same value by construction. 
For each input value, the darker filled curve shows the likelihood obtained using the full covariance $\boldsymbol{C}_{\rm GG}$, while the lighter filled curve of the same color uses Poisson errors. The difference between the two is therefore entirely due to the error budget: the covariance-based constraints are systematically and significantly broader than their Poisson counterparts, reflecting the combined effect of cosmic variance and off-diagonal bin-to-bin correlations that Poisson errors neglect entirely.

Fig.~\ref{fig:sigma_ratio_heatmap} presents the ratio $\sigma^{\mminq}_{\rm cov} / \sigma^{\mminq}_{\rm Poisson}$ for the marginalized $\mminq$ constraint from the full joint QG$+$GG fit across the grid of 30 mass pairs. The general trend is clear: the Poisson approximation underestimates the true error more severely at higher halo masses, for both $\mminq$ and $\mming$. At low masses ($\mminq = 11.5$, $\mming = 10.5$), the ratio is modest ($\sigma^{\mminq}_{\rm cov}/\sigma^{\mminq}_{\rm Poisson} \approx 1.4$--$1.5$), while at high masses ($\mminq \geq 12.1$, $\mming = 10.9$) it reaches $\approx$2.4--$2.7$. However, at the highest quasar halo masses ($\log\mminq/\msun \geq 12.3$, top two rows of Fig.~\ref{fig:sigma_ratio_heatmap}), the ratio does not grow monotonically. This is driven by two boundary effects of the precomputed grids: the mock covariance was generated only on $\log\mminq/\msun \in [11.5, 12.5]$, so inferences at higher input masses fall back to the $\mminq = 12.5$ covariance rather than the true mass-dependent covariance; and the model $\chi(r_p)$ grid extends only to $\log\mminq/\msun = 13.0$, truncating the posterior at this upper boundary. Because the covariance posterior is broader than the Poisson posterior, it is clipped more aggressively by this upper-boundary truncation, so $\sigma^{\mminq}_{\rm cov}$ shrinks faster than $\sigma^{\mminq}_{\rm Poisson}$ as $\mminq$ approaches the grid edge. The ratio therefore stops growing — and in some cells decreases — at the top of the heatmap, instead of continuing to increase with $\mminq$ as it does at lower masses.

This mass-dependent trend can be understood from the off-diagonal elements of the correlation matrix becoming more prominent at higher halo masses. As illustrated in Figs.~\ref{fig:cov_cross} and \ref{fig:cov_auto}, the positive off-diagonal correlations arise because large-scale density fluctuations boost galaxy pair counts at all separations simultaneously: an overdense region increases the number of pairs in every $r_p$ bin, causing the correlation function values to rise and fall together across bins. For more massive (i.e., more highly biased) halos, these correlated fluctuations are amplified because the clustering response to the underlying density field is stronger \citep{Valageas2011, Mo1996}. This is consistent with the trend seen in Fig.~\ref{fig:sigma_ratio_heatmap}, where the ratio increases with both $\mminq$ and $\mming$. 


\subsection{Inference test}
\label{sec:inference_test}

The mass-dependent comparison in Figs.~\ref{fig:likelihood_Mgal} and \ref{fig:sigma_ratio_heatmap} shows that Poisson errors underestimate the width of the halo mass constraints by a factor of $\approx 1.5$--$2.7$. We now verify that the narrower Poisson credible intervals fail to contain the true parameters at the expected rate, using a coverage test \citep[see appendix~A of][]{Hennawi2025}, which checks whether the resulting credible intervals contain the true parameters at the expected rate.  We perform a coverage test using 1000 mock realizations generated at the fiducial input masses $(\log\mming/\msun, \log\mminq/\msun) = (10.6, 12.2)$. For each realization, we run two independent MCMC inferences, one with the full mock covariance matrix and one with diagonal Poisson errors, and record the log-posterior probability evaluated at the MCMC samples, $\ln P(\theta_i)$, and at the true input parameters, $\ln P(\theta_{\rm true})$.

For a given credible level $\alpha$, we define the $\alpha$-contour as the iso-probability surface enclosing fraction $\alpha$ of the posterior mass. We then measure the coverage $C(\alpha)$: the fraction of mock realizations for which the true parameters fall within the $\alpha$-contour. A posterior with perfect coverage satisfies $C(\alpha) = \alpha$, meaning that the $\alpha$-credible region contains the true parameters in exactly fraction $\alpha$ of realizations. If $C(\alpha) < \alpha$, the credible regions are too narrow (overconfident); if $C(\alpha) > \alpha$, they are too wide (underconfident).

\begin{figure}
\centering
    \includegraphics[width=\columnwidth]{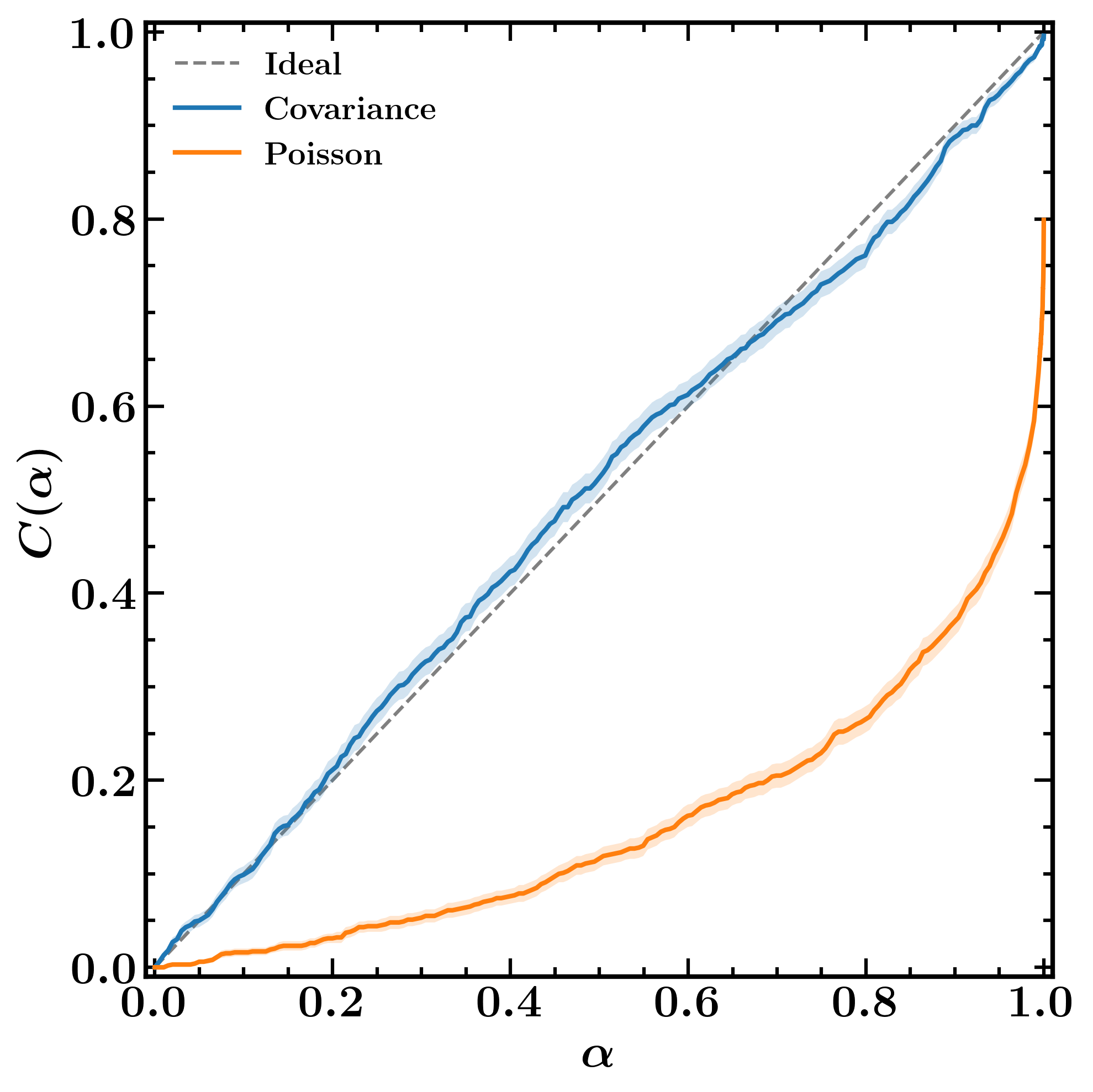}
    \caption{Coverage test comparing the covariance-based and Poisson likelihoods across 1000 mock realizations at the fiducial input masses $(\log\mming/\msun, \log\mminq/\msun) = (10.6, 12.2)$. The coverage $C(\alpha)$ gives the fraction of realizations for which the true parameters fall within the $\alpha$-credible region of the posterior. Shaded bands show $1\sigma$ binomial uncertainties. The covariance likelihood (blue) follows the ideal diagonal, confirming that the full mock covariance correctly captures the scatter across realizations. The Poisson likelihood (orange) is severely overconfident: at $\alpha=0.68$, the true parameters are contained in only $\approx$20\% of realizations instead of the expected 68\%, because the Poisson approximation ignores cosmic variance and inter-bin correlations.
    }
    \label{fig:coverage_test}
\end{figure}

Fig.~\ref{fig:coverage_test} shows the resulting coverage curves $C(\alpha)$ for both likelihoods, with $1\sigma$ binomial confidence bands from the 1000 realizations. The covariance-based posterior (blue) closely follows the ideal diagonal $C(\alpha)=\alpha$, confirming that the full mock covariance matrix, which encodes cosmic variance, inter-bin correlations, and the non-independent nature of galaxy pairs, produces credible intervals with the correct coverage. This also validates the Gaussian likelihood assumption: despite the discrete nature of pair counts at small separations, the Gaussian likelihood combined with the full covariance matrix produces well-calibrated credible intervals. In contrast, the Poisson posterior (orange) falls dramatically below the diagonal: at $\alpha = 0.68$ (the nominal $1\sigma$ level), the true parameters are contained in only $\approx$20\% of the Poisson credible regions. This is a direct consequence of the Poisson approximation ignoring cosmic variance and treating all radial bins as statistically independent, which artificially shrinks the posterior and leads to severe overconfidence in the inferred halo masses.

We note that a fully rigorous inference test \citep{SelentinHeavens2016, Talts2020, Hennawi2025}
would draw the true parameters from the prior for each mock realization, requiring a separate set of mock catalogs for every $(\mming, \mminq)$ draw. Since our 1000 realizations share a single set of input masses, our test measures the \textit{conditional} coverage at one point in parameter space. Nevertheless, this is sufficient to demonstrate that the Poisson likelihood systematically underestimates the true uncertainty, whereas the covariance-based likelihood does not.

 \subsection{Effect of Separate Central and Satellite Mass Thresholds}
\label{sec:split_hod}

The minimum halo mass inference framework presented in \S\S\ref{sec:likelihood}--\ref{sec:inference_test} adopts a single galaxy mass threshold $\mming$ shared by both central and satellite \oiii-emitters. We now relax this assumption and investigate how the cross- and auto-correlation functions, and the inference of the underlying halo masses, change when centrals and satellites are allowed to populate halos above different mass thresholds.

\subsubsection{Physical motivation}

The single-threshold HOD assumes that the probability of a galaxy having a given luminosity depends only on its (sub)halo peak mass $M_{\rm peak}$, regardless of whether the galaxy is a central or a satellite.
In practice, however, the luminosity distribution $L(M_{\rm peak})$ is unlikely to be the same for both populations.
Several environmental processes systematically alter the properties of satellite galaxies after infall, including strangulation of the hot gas supply \citep{Larson1980, Balogh2000}, ram pressure stripping of the cold gas reservoir \citep{Gunn1972, Abadi1999}, and tidal stripping and harassment from the host potential and neighboring subhalos \citep{Moore1996, Read2006}.
These effects could either suppress or, in some cases, temporarily boost (e.g., via ram-pressure-induced starbursts) the \oiii\ luminosity of satellites relative to centrals of the same $M_{\rm peak}$.

Since our galaxy sample is selected via an \oiii\ flux limit, any systematic difference in $L(M_{\rm peak})$ between centrals and satellites translates directly into different effective mass thresholds for the two populations, which in turn changes the satellite fraction of the observed sample. For instance, if satellite luminosities are systematically suppressed at fixed $M_{\rm peak}$, the flux-limited sample preferentially selects satellites from more massive subhalos, yielding an effective $M_{\rm min,sat} > M_{\rm min,cen}$ and a smaller satellite fraction. Conversely, if environmental interactions temporarily boost satellite \oiii\ emission, the effective satellite threshold could be lower than the central threshold and the satellite fraction correspondingly larger.

For QSOs, the integrated black hole mass is relatively robust to the central--satellite distinction, since much of the black hole growth likely occurred while the galaxy was still a central. The instantaneous accretion rate that powers an observed QSO phase is more vulnerable: gas stripping in satellites depletes the cold-gas reservoir that fuels the central black hole, so a single $\mminq$ would over-count QSOs from the satellite population. We therefore restrict QSOs to centrals throughout this section (\S\ref{sec:split_hod_mocks}); a fully general treatment would also split $\mminq$. For emission-line luminosities, which respond to instantaneous star formation rates, the central--satellite distinction is similarly consequential, motivating the split $\mmincen$/$\mminsat$ treatment that follows.

\subsubsection{Model setup and mock generation}
\label{sec:split_hod_mocks}

We generalize the step-function HOD adopted throughout this work by introducing separate halo-mass thresholds for central and satellite \oiii-emitters, $\mmincen$ and $\mminsat$, while retaining a single threshold $\mminq$ for QSOs. Each subhalo is classified using its rank in the halo catalog: rank $=0$ identifies a central, while rank $>0$ identifies a satellite. A halo or subhalo with $M_{\rm peak}$ above its corresponding threshold hosts exactly one tracer, with no scatter. QSOs are assumed to populate only central subhalos (rank~$=0$), with their position taken as the most bound particle of the host. To remove numerically unresolved objects we additionally require that every subhalo contains at least 20 bound dark-matter particles ($N_{\rm bound} \geq 20$), consistent with the resolution cut described in \S\ref{sec:cov}.

The remainder of the mock construction follows the procedure in \S\ref{sec:mock_construction}. We use the same FLAMINGO-10k snapshot at $z_{\rm snap} = 6.14$, the same 25 ASPIRE pointings with their selection functions, the same line-of-sight integration scale $\pi_{\max} = 7\,\mpch$, and the same set of $r_p$ bins. The only change is in step (i) in \S\ref{sec:mock_construction}: instead of a single galaxy mass cut, we now select centrals from subhalos with rank~$=0$ and $M_{\rm peak} > {\mmincen}\,\msun$, and satellites from subhalos with rank~$>0$ and $M_{\rm peak} > {\mminsat}\,\msun$. Selection function and down-sampling to the \citet{Matthee2023} number density are then applied as in \S\ref{sec:mock_construction}.

For the inference described in \S\ref{sec:split_hod_inference} we precompute $\chi_{\rm GG}(r_p)$ and $\chi_{\rm QG}(r_p)$ on a grid of $(\mminq, \mmincen, \mminsat)$. The grid spans $\log\mminq/\msun \in [11.25, 12.75]$ in steps of $0.25$~dex, $\log\mmincen/\msun \in [10.5, 11.5]$ in steps of $0.1$~dex, and $\log\mminsat/\msun \in [10.5, 12.5]$ in steps of $0.1$~dex. Likelihood evaluations during MCMC use linear interpolation on this grid. 

\begin{figure*}
\centering
    \includegraphics[width=\textwidth]{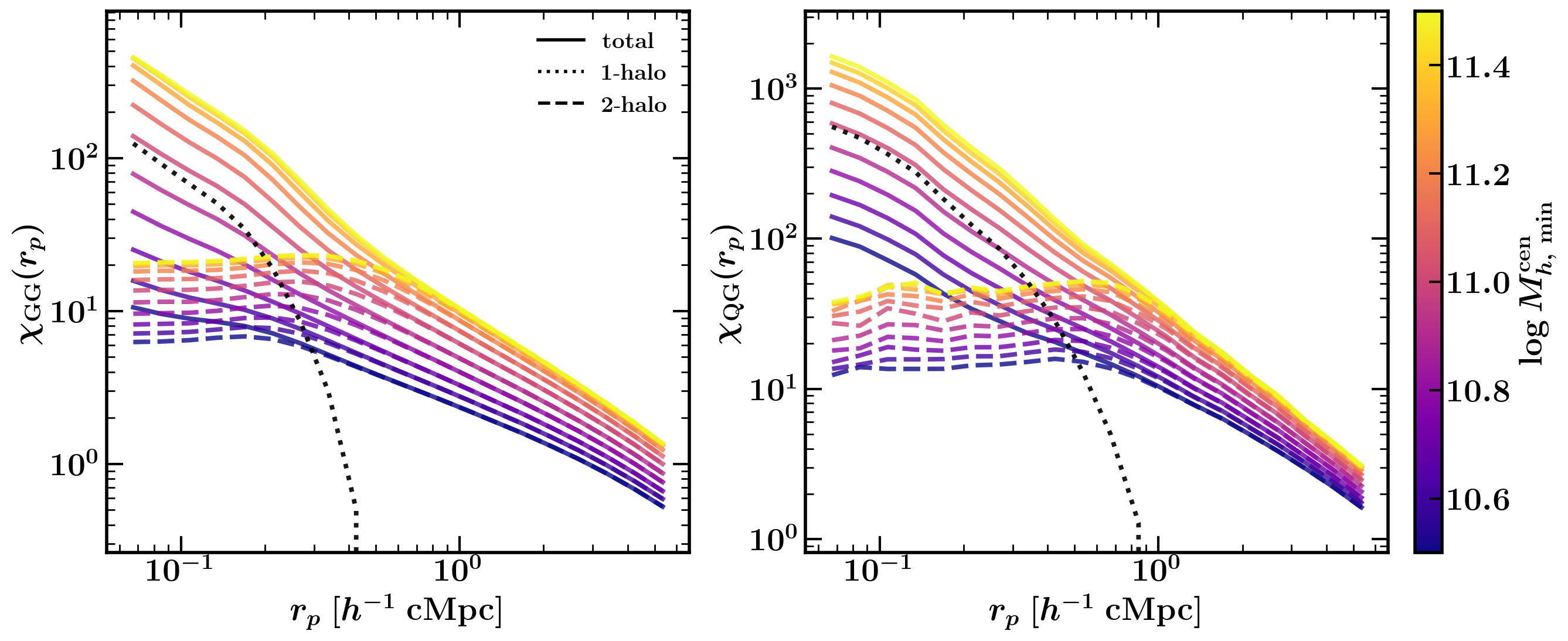}
\caption{Sensitivity of the volume-averaged correlation functions to the minimum mass of the central host halos for \oiii-emitters $\log\mmincen/\msun$, at fixed $\log\mminq/\msun = 12.20$ and $\log\mminsat/\msun = 10.60$. \textbf{\textit{Left:}} \oiii\ auto-correlation $\chi_{\rm GG}(r_p)$. \textbf{\textit{Right:}} quasar--\oiii\ cross-correlation $\chi_{\rm QG}(r_p)$. In both panels, solid colored curves show the total correlation function, dotted colored curves show the 2-halo term, and the single black dashed curve shows the 1-halo term, which is essentially independent of $\mmincen$ at fixed $\mminsat$. Increasing $\mmincen$ selects centrals from more massive halos, raising the mean galaxy bias and boosting the 2-halo and total amplitudes; the 1-halo curves remain unchanged because they are set by $\mminsat$, which is held fixed.}
\label{fig:chi_sweep_Mcen}
\end{figure*}

\begin{figure*}
\centering
    \includegraphics[width=\textwidth]{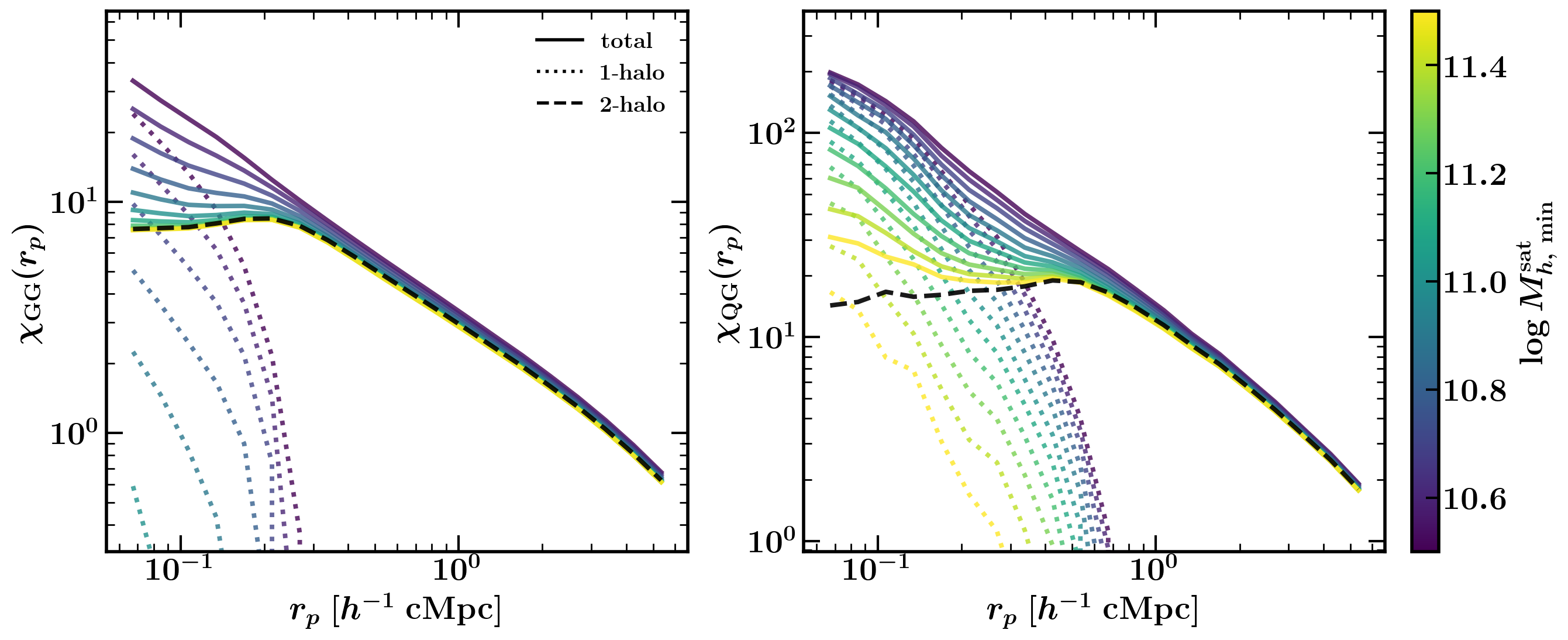}
\caption{Same as Fig.~\ref{fig:chi_sweep_Mcen}, but varying the satellite halo-mass threshold $\log\mminsat/\msun$ at fixed $\log\mminq/\msun = 12.20$ and $\log\mmincen/\msun = 10.70$. \textbf{\textit{Left:}} $\chi_{\rm GG}(r_p)$. \textbf{\textit{Right:}} $\chi_{\rm QG}(r_p)$. In both panels, solid colored curves show the total correlation function, dotted colored curves show the 1-halo term, and the single black dashed curve shows the 2-halo term, which is set by the (fixed) central population. The 1-halo curves drop steeply at small $r_p$ as $\mminsat$ increases and strips satellites out of QSO and galaxy host halos, while the large-$r_p$ 2-halo signal is essentially unchanged. Comparing to Fig.~\ref{fig:chi_sweep_Mcen}, the small-scale 1-halo regime carries essentially all of the information about $\mminsat$ in this model, while the large-scale 2-halo amplitude is controlled by $\mmincen$.}
\label{fig:chi_sweep_Msat}
\end{figure*}

Figs.~\ref{fig:chi_sweep_Mcen} and \ref{fig:chi_sweep_Msat} illustrate how the two galaxy thresholds shape the volume-averaged auto- and cross-correlation functions. Each figure shows the \oiii\ auto-correlation $\chi_{\rm GG}$ on the left and the QSO--\oiii\ cross-correlation $\chi_{\rm QG}$ on the right, with the swept threshold encoded by the colormap; in each case we also decompose the signal into 1-halo and 2-halo contributions. We note that the 1-halo terms (in Figs.~\ref{fig:chi_sweep_Mcen} and \ref{fig:chi_sweep_Msat}) remain finite at small $r_p$ rather than dropping to zero as halo exclusion would imply for the 3D correlation function $\xi(r)$; this is a projection effect of the volume-averaged $\chi(r_p)$ definition (Eq.~\ref{eq:volavg_corr}), which integrates over the cylindrical shell with $|\pi| \leq \pi_{\max}$ and therefore picks up contributions from pairs with larger 3D separations even at small $r_p$.

\paragraph*{Varying the central mass threshold.}
When $\mmincen$ is varied with $\mminsat$ held fixed (Fig.~\ref{fig:chi_sweep_Mcen}):
\begin{enumerate}
    \item The one-halo term (single black dashed line) is essentially unchanged across all models in both $\chi_{\rm GG}$ and $\chi_{\rm QG}$, since the satellite population in galaxy host halos and in quasar host halos depends only on $\mminsat$, which is held fixed.

    \item The two-halo term (colored dotted lines) increases with $\mmincen$ in both $\chi_{\rm GG}$ and $\chi_{\rm QG}$, as expected: selecting centrals from more massive halos increases the mean galaxy bias.

    \item The total auto- and cross-correlations (solid colored lines) increase with $\mmincen$ beyond the two-halo contribution alone. At fixed total galaxy number density, increasing $\mmincen$ reduces the number of centrals while the satellite count remains fixed, raising the satellite fraction from $f_{\rm sat} \approx 5\%$ at $\log\mmincen/\msun = 10.5$ to $\approx 87\%$ at $\log\mmincen/\msun = 11.5$. The higher satellite fraction increases the relative weight of the one-halo term in the total signal of both $\chi_{\rm GG}$ and $\chi_{\rm QG}$.
\end{enumerate}

\paragraph*{Varying the satellite mass threshold.}
When $\mminsat$ is varied with $\mmincen$ held fixed (Fig.~\ref{fig:chi_sweep_Msat}):
\begin{enumerate}
    \item The two-halo term is insensitive to $\mminsat$ in both the auto- and cross-correlation. At $r_p \gtrsim 1$~cMpc~$h^{-1}$, all colored curves converge to the single black dashed 2-halo line in both $\chi_{\rm GG}$ (left) and $\chi_{\rm QG}$ (right). The two-halo clustering amplitude depends on the total galaxy bias, which is dominated by the more numerous centrals; since the central population is identical across all models, the large-scale signal is nearly unchanged.

    \item The one-halo term depends strongly on $\mminsat$ in both $\chi_{\rm GG}$ and $\chi_{\rm QG}$. At $r_p \lesssim 0.5$~cMpc~$h^{-1}$, the one-halo signal arises from satellites sharing a host halo with another tracer: another galaxy in the case of $\chi_{\rm GG}$, and a quasar in the case of $\chi_{\rm QG}$. Lowering $\mminsat$ adds more satellites to these halos and boosts the one-halo pair count. The effect is more pronounced in $\chi_{\rm QG}$ because quasar halos are massive ($M_h \geq 10^{12.2}\,\msun$) and host many satellites: the satellite fraction within quasar halos grows faster than the global mean galaxy density, so $\chi_{\rm QG}^{\rm 1h}$ has a larger dynamic range in $\mminsat$ than $\chi_{\rm GG}^{\rm 1h}$.
\end{enumerate}

The complementary roles of $\mmincen$ (large-scale, 2-halo) and $\mminsat$ (small-scale, 1-halo) are the physical reason the two thresholds are independently constrainable when both auto- and cross-correlation functions are fit jointly.

\subsubsection{Inference with the full mock covariance matrix}
\label{sec:split_hod_inference}

The likelihood structure follows \S\ref{sec:likelihood} (Eqs.~\ref{eq:total_likelihood}--\ref{eq:auto_likelihood}), with the model means $\chi_{\rm GG}^{\rm model}$ and $\chi_{\rm QG}^{\rm model}$ evaluated as functions of the three free parameters $(\mminq, \mmincen, \mminsat)$ via linear interpolation on the precomputed grid above. In principle the covariance matrices should also be made mass-dependent, following the construction in \S\ref{sec:cov} where $\boldsymbol{C}_{\rm GG}(\mming)$ and $\boldsymbol{C}_{\rm QG}(\mminq, \mming)$ are computed on the two-parameter grid, with the natural extension here being $\boldsymbol{C}_{\rm GG}(\mmincen, \mminsat)$ and $\boldsymbol{C}_{\rm QG}(\mminq, \mmincen, \mminsat)$ on the three-parameter grid. However, since our goal in this section is to demonstrate the effect of separating the central and satellite thresholds rather than to fully propagate their impact on the error budget, we therefore adopt a single covariance matrix for each of $\chi_{\rm GG}$ and $\chi_{\rm QG}$, computed once at the fiducial truth $\log[(\mminq, \mmincen, \mminsat)/\msun] = (12.20, 10.70, 10.60)$, and hold it fixed during MCMC sampling. The Poisson comparison case follows \S\ref{sec:likelihood} exactly: the per-bin Poisson confidence interval is computed from the raw $D_{\rm Q}D_{\rm G}$ and $D_{\rm G}D_{\rm G}$ pair counts of the realization being fit, and propagated through the Landy--Szalay and Davis--Peebles estimators (Eqs.~\ref{eq:auto}--\ref{eq:cross}) to give per-bin errors on $\chi_{\rm GG}$ and $\chi_{\rm QG}$.

The $(\mmincen, \mminsat)$ samples are mapped to derived satellite and central fractions $f_{\rm sat}$ and $f_{\rm cen} = 1 - f_{\rm sat}$, computed by counting the centrals (rank~$=0$) above $\mmincen$ and the satellites (rank~$>0$) above $\mminsat$ in the FLAMINGO-10k subhalo catalog with the same $N_{\rm bound} \geq 20$ resolution cut.

\begin{figure}
  \centering
  \includegraphics[width=\columnwidth]{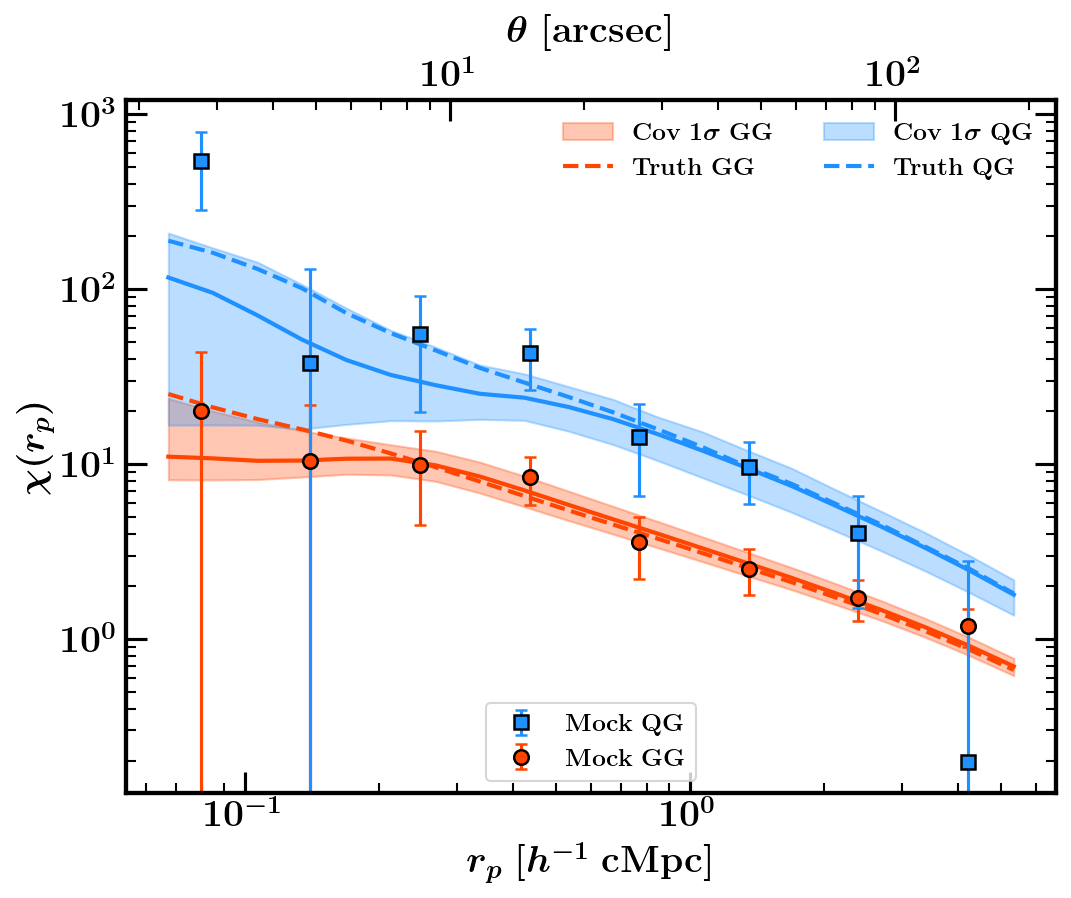}
  \caption{Best-fit $\chi_{\rm GG}(r_p)$ (red) and $\chi_{\rm QG}(r_p)$ (blue) for a single mock realization at the fiducial truth $\log[(\mminq, \mmincen, \mminsat)/\msun] = (12.20, 10.70, 10.60)$. Errorbars are computed from the diagonal of the full mock covariance at the fiducial model. The dashed red curve is the truth at the input parameters. Solid lines and shaded bands show the posterior median and the $16$--$84$\% range of the model.}
  \label{fig:split_hod_bestfit}
  \end{figure}

  \begin{figure}
  \centering
  \includegraphics[width=\columnwidth]{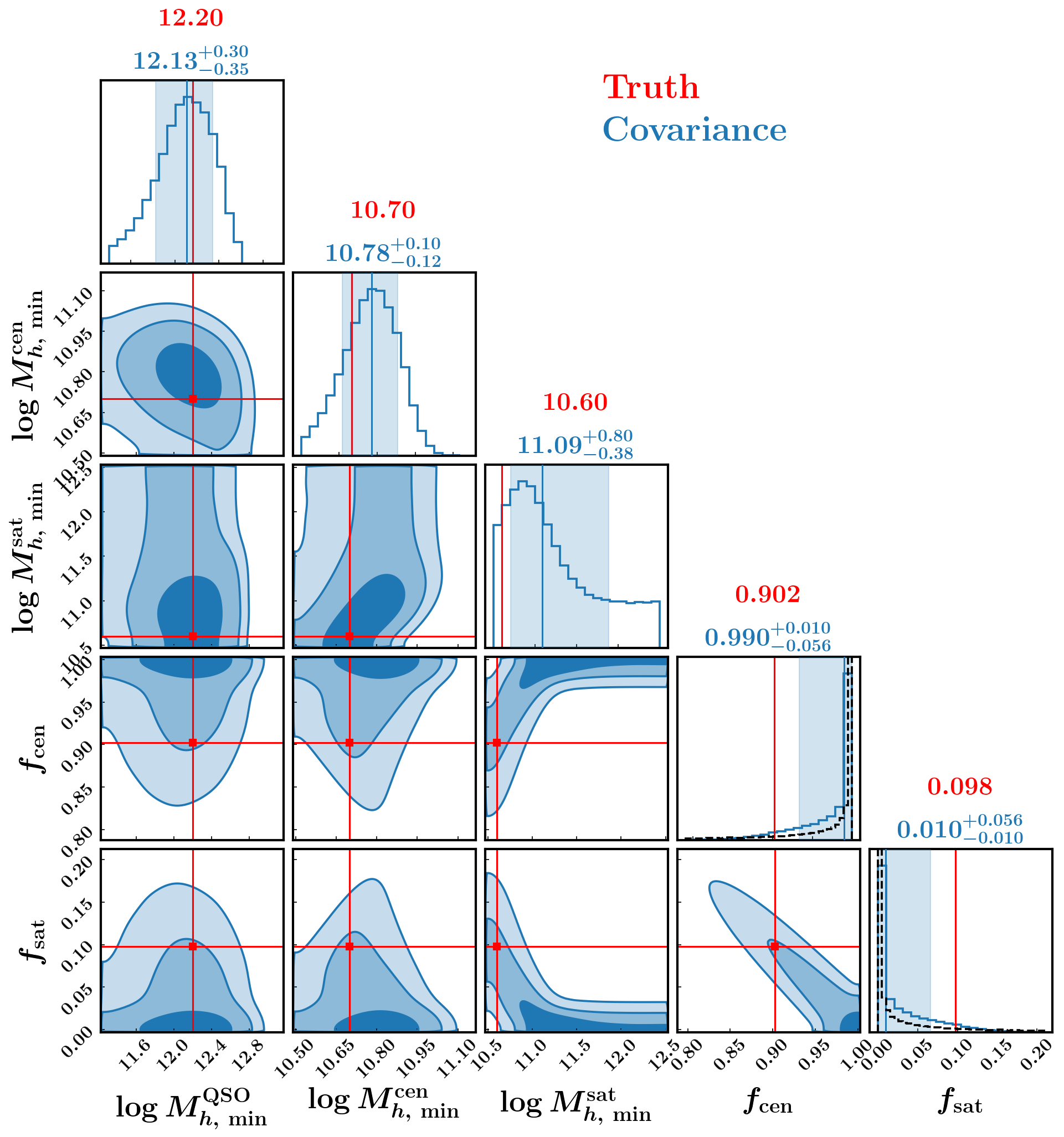}
  \caption{Joint posterior in the five parameters $(\log[(\mminq, \mmincen, \mminsat)/\msun], f_{\rm cen}, f_{\rm sat})$ for the same realization as Fig.~\ref{fig:split_hod_bestfit}. Contours show the posterior using the full mock covariance, enclosing 1, 2, and $3\sigma$ Gaussian-equivalent posterior mass. Red lines mark the truth. The lower two parameters $(f_{\rm cen}, f_{\rm sat})$ are derived deterministically from $\log[(\mmincen, \mminsat)/\msun]$ via the FLAMINGO subhalo
  halo-mass functions and carry no independent information. The dashed black curves on the $f_{\rm cen}$ and $f_{\rm sat}$ 1D diagonals show the induced prior on these fractions, obtained by mapping the flat $\log[(\mmincen, \mminsat)/\msun]$ priors.}
  \label{fig:split_hod_corner}
  \end{figure}

Fig.~\ref{fig:split_hod_bestfit} shows the inference result for one random realization, with the covariance-based posterior recovering the truth model within the bands. Fig.~\ref{fig:split_hod_corner} shows the corresponding 5-parameter posterior; the covariance contours enclose the truth at the $2\sigma$ level in all three free parameters. 

Unlike the mass thresholds, $f_{\rm cen}$ and $f_{\rm sat}$ are derived rather than free parameters, and a flat prior in $\log\mmincen$ and $\log\mminsat$ does \textit{not} translate to a flat prior on $f_{\rm cen}$ and $f_{\rm sat}$. The mapping $(\mmincen, \mminsat) \to (f_{\rm cen}, f_{\rm sat})$ has a non-trivial Jacobian: because the satellite halo-mass function is steep, $n_{\rm sat}$ falls rapidly with $\mminsat$, so that for $\log\mminsat/\msun \gtrsim 11$ it becomes negligible compared to $n_{\rm cen}$ and drives $f_{\rm sat} \to 0$, $f_{\rm cen} \to 1$. Most of the flat prior on $\mminsat \in U(10.5, 12.5)$ lies in this regime. The induced prior on $f_{\rm cen}$ is therefore strongly skewed toward 1 (and the prior on $f_{\rm sat}$ toward 0), as shown by the dashed black curves on the $f_{\rm cen}$ and $f_{\rm sat}$ panels of Fig.~\ref{fig:split_hod_corner}. We keep the priors flat in log-mass and accept this Jacobian, because the experiment probes mass thresholds rather than fractions; flattening the $f_{\rm cen}$ prior would impose a non-trivial prior on the masses we are trying to infer. The Jacobian also explains why the recovered posteriors on $f_{\rm cen}$ sit close to 1: the fiducial truth $\log\mminsat/\msun = 10.60$ lies near the lower edge of the $\mminsat$ prior, in a regime where the prior already pulls $f_{\rm cen}$ toward 1.

The marginalized posterior on $\mminq$ is essentially unchanged between the single-threshold analysis of \S\ref{sec:mmin} and the analysis with separate central and satellite thresholds here, indicating that the QSO halo-mass inference is robust to the central--satellite separation in the galaxy sample.

\subsubsection{Implications for the single-threshold HOD}

These results demonstrate that $\mmincen$ and $\mminsat$ affect the volume-averaged correlation functions in distinct ways. Varying $\mmincen$ changes the mean galaxy bias, which controls the two-halo amplitude at large separations, and at fixed total galaxy number density also increases the satellite fraction, boosting the relative weight of the one-halo term. Varying $\mminsat$ primarily affects the one-halo term at small projected separations ($r_p \lesssim 0.5~\mpch$), while leaving the two-halo regime essentially unchanged.

This has two practical consequences for the analysis presented in this work:
\begin{enumerate}
    \item The standard single-threshold HOD is a good approximation in the two-halo regime that dominates the total signal and drives the halo mass inference. The two-halo term is controlled by the central population, which is the same in both the single-threshold and split models when $\mmincen = \mming$.

    \item At small scales where one-halo contributions become significant, the single-threshold model could introduce a systematic bias if the true central and satellite mass thresholds differ substantially. Several mock data points fall within the one-halo radial range ($r_p \lesssim 0.5$~cMpc~$h^{-1}$), and even though these data points do not dominate the fit, they do contribute to the total likelihood. If environmental processes have preferentially suppressed (or boosted) satellite \oiii\ luminosities, the effective satellite mass threshold would differ from the central threshold, introducing a model misspecification that is not captured by our current framework.
\end{enumerate}

We flag this as a systematic uncertainty that future analyses with larger datasets and deeper spectroscopy should address. In principle, the one-halo regime offers a direct handle on the satellite fraction and could be used to jointly constrain $\mmincen$ and $\mminsat$ (and hence also the satellite fraction), provided sufficient signal-to-noise at small separations. At $z \approx 6$, the relatively short time since satellite infall may limit the magnitude of environmental effects, but they cannot be excluded \textit{a priori}, particularly for emission-line selected samples where even modest changes in the instantaneous star formation rate translate directly into changes in \oiii\ detectability.

\section{Discussion and Conclusions}
\label{sec:conclusion}

We have developed a framework for robust high-redshift clustering inference by constructing realistic mock observations from the FLAMINGO-10k (F10k) dark-matter-only simulation. By generating 1000 mock realizations matched to the ASPIRE setup of 25 independent \textit{JWST}/NIRCam WFSS pointings around $z\approx 6.6$ quasars (including the field-by-field spatial coverage and sensitivity maps), we construct mock covariance matrices for the \OIII-emitter auto-correlation function $\chi_{\rm GG}(r_p)$ and the quasar--\OIII-emitter cross-correlation function $\chi_{\rm QG}(r_p)$. These covariance matrices are computed on a grid of minimum halo masses $(\mminq, \mming)$, capturing the mass dependence of cosmic variance. We validate the inference framework through coverage tests, demonstrating that the covariance-based credible intervals have the correct statistical properties, and quantify the impact of adopting Poisson pair-count errors versus the full mock covariance matrices on inferred halo masses and correlation lengths. Our main results are as follows:

\begin{enumerate}

\item Poisson errors underestimate the true uncertainty by factors of $\approx$1.5--2.7. By comparing the $1\sigma$ constraints on $\mminq$ from the joint QG$+$GG fit using the full covariance versus diagonal Poisson errors across a grid of 30 mass configurations, we find a systematic ratio $\sigma^{\mminq}_{\rm cov}/\sigma^{\mminq}_{\rm Poisson}$ that increases with both $\mminq$ and $\mming$ (Fig.~\ref{fig:sigma_ratio_heatmap}). At fiducial ASPIRE masses ($\log(\mminq/\msun) = 12.2$, $\log(\mming/\msun) = 10.6$), the Poisson posterior on the minimum halo masses is $\approx 1.8\times$ narrower than the covariance-based posterior (Fig.~\ref{fig:mock_inference}).

\item The covariance structure is mass dependent. The off-diagonal elements of the covariance matrix become more prominent at higher halo masses (Figs.~\ref{fig:cov_cross} and \ref{fig:cov_auto}), because more massive, more highly biased tracers cluster more strongly, producing larger correlations and correspondingly larger covariance.
This means that using a fixed fiducial covariance matrix, while preferable to Poisson errors, may still not capture the full mass dependence of the error budget. We therefore advocate for mass-dependent covariance matrices in the likelihood, as implemented in this work.


\item The inference test confirms that Poisson credible intervals are severely overconfident. Across 1000 mock realizations at the fiducial masses, the covariance-based $1\sigma$ credible region contains the true parameters in $\approx$65\% of realizations, consistent with the expected 68\% within statistical uncertainty, while the Poisson $1\sigma$ region contains the true parameters in only $\approx$20\% of realizations (Fig.~\ref{fig:coverage_test}).

\item The QSO halo-mass inference is robust to the central--satellite treatment of the galaxy sample. The marginalized posterior on $\mminq$ is essentially unchanged between the single-threshold analysis (\S\ref{sec:mmin}) and the analysis with separate central and satellite thresholds (\S\ref{sec:split_hod}), indicating that the inferred $\mminq$ does not depend on the detailed treatment of the 1-halo term.



\end{enumerate}

Our results have important implications for existing and future high-redshift clustering studies. The commonly adopted Poisson error bars on the two-point correlation function at $z \gtrsim 6$ should be regarded as lower bounds on the true uncertainty. Studies that have reported halo mass constraints without full simulation-based covariance matrices \citep[e.g.,][]{Lee2006, BaroneNugent2014, Harikane2016, Harikane2022, GarciaVergara2017, GarciaVergara2019, Arita2023, Eilers2024, Schindler2026, Shuntov2025, Lin2025_lf}
likely underestimate the true error bars on inferred halo masses by factors of $\approx$1.5--3, depending on the halo mass and survey geometry. This does not necessarily invalidate the central values of previous measurements, but it does mean that reported confidence intervals are too narrow, and that apparent tensions between studies may be less significant than they appear. Robust covariance-based methodology is well established at low-to-intermediate redshifts (see \S\ref{sec:introduction} and references therein), but has not previously been applied to clustering at $z \gtrsim 6$.

The framework presented here is readily applicable to other \textit{JWST} WFSS programs targeting emission-line galaxies. In particular, the COSMOS-3D survey \citep{Kakiichi2024} covers $\approx$0.3~deg$^2$ with NIRCam/WFSS, providing a large sample of emission-line galaxies \citep{Meyer2025}. The galaxy auto-correlation function measured from this blank-field survey can constrain the host halo mass and duty cycle of emission-line galaxies, and the mock covariance framework developed here can be adapted to the COSMOS-3D survey geometry to ensure robust error estimates. Other WFSS programs, including pure-parallel surveys such as SAPPHIRES \citep{Sun2025_SAPPHIRES}, will further expand the sample of emission-line galaxies available for clustering analyses across multiple redshifts and tracers. Even in larger contiguous-mosaic surveys such as COSMOS-3D ($\approx 0.3$~deg$^2$, $\approx 4.5\times$ the total ASPIRE area and sampled in a single connected volume rather than 25 independent pointings), cosmic variance does not become negligible: applying our framework to mock COSMOS-3D \OIII-emitter auto-correlation measurements, we find that Poisson errors still underestimate the true uncertainty of the halo mass by a factor of $\approx 2$ (Huang et al., in prep.). Proper covariance modeling therefore remains essential across the range of current and near-future JWST/WFSS surveys to avoid overconfident constraints on the galaxy--halo connection.


\section*{Acknowledgements}
We acknowledge helpful conversations with the ENIGMA group at UC Santa Barbara and Leiden University. JH is grateful to the discussions with Feige Wang, Shane Bechtel, Daming Yang, Ben Wang, Bingcheng Jin, Xiaojing Lin, and Romain Meyer. JH and JFH acknowledges support from the National Science Foundation under Grant No. 2307180.
JFH also acknowledges support for program XXX provided by NASA through a grant from the Space Telescope Science Institute, which is operated by the Association of Universities for Research in Astronomy, Inc., under NASA contract NAS 5-03127.
EP acknowledges support provided by NASA through the NASA Hubble Fellowship grant HST-HF2-51586.001-A awarded by the Space Telescope Science Institute, which is operated by the Association of Universities for Research in Astronomy, Incorporated, under NASA contract NAS5-26555.

\section*{Data Availability}

The data underlying the findings of this study, together with the analysis code used to produce the results, are available from the corresponding author upon request.

\newpage


\bibliographystyle{mnras}
\bibliography{citations} 




\appendix
\newpage

\section{Inference with the mean mock correlation function}
\label{appendix:mean_mock}

As a complement to the single-realization inference shown in Fig.~\ref{fig:mock_inference}, we repeat the MCMC analysis using the mean of all 1000 mock realizations as the ``observed'' data vector. Because the mean mock correlation function converges to the true halo model prediction, this isolates the effect of the error model (covariance vs.\ Poisson) on the posterior width without the confounding effect of noise fluctuations in any particular realization.

\begin{figure*}
\centering
    \includegraphics[width=0.48\textwidth]{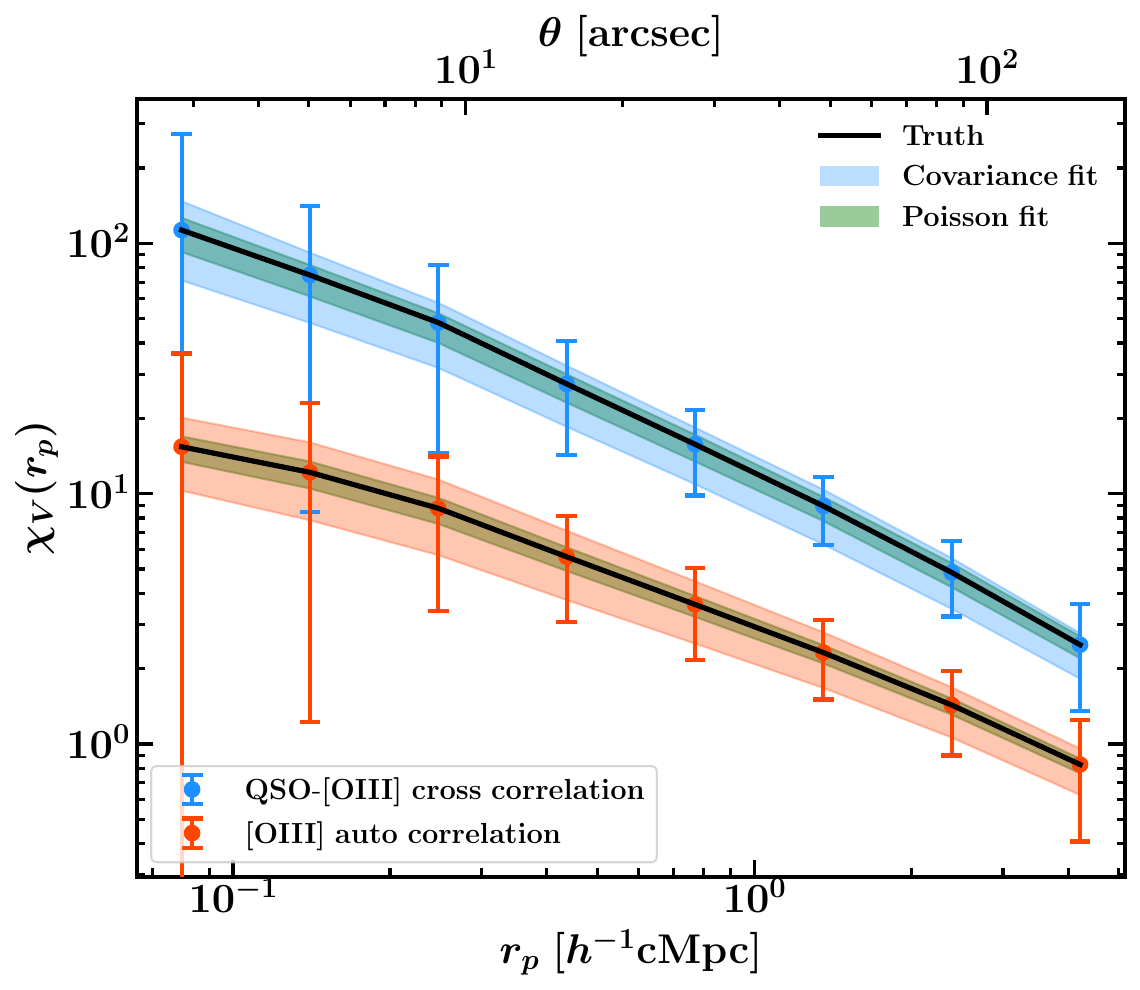}
    \hfill
    \includegraphics[width=0.48\textwidth]{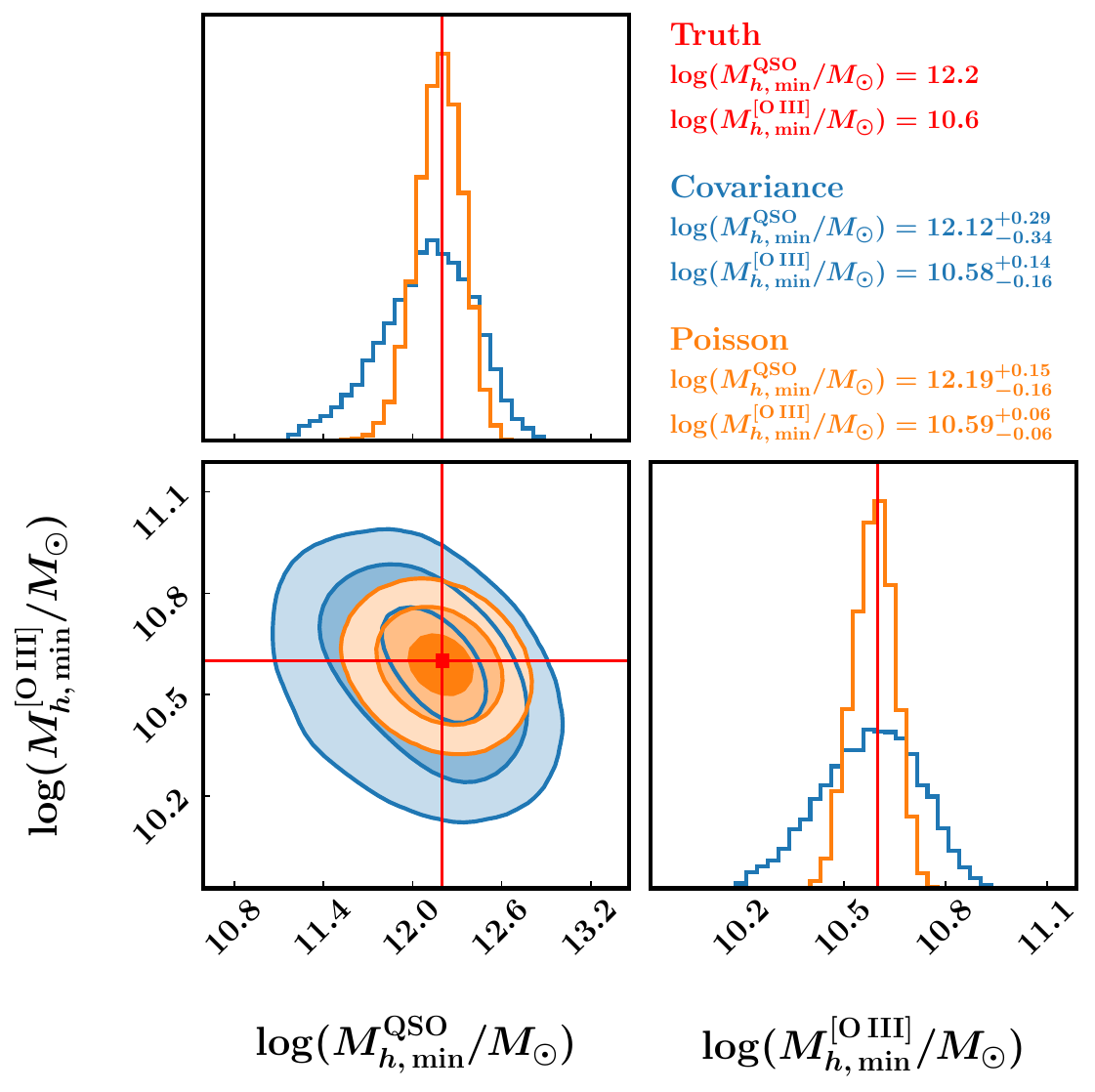}
    \caption{Same as Fig.~\ref{fig:mock_inference}, but using the mean of 1000 mock realizations as the observed data. \textbf{\textit{Left:}} Since the mean converges to the true model, the data points lie on the black truth curves by construction. Only the $1\sigma$ posterior bands are shown (no median curves). The covariance fit bands (blue/red shaded) remain broader than the Poisson fit bands (green shaded), confirming that the broadening is a systematic effect of the error model rather than noise in any single realization. \textbf{\textit{Right:}} Both posteriors peak at the true input masses, but the covariance-based posterior (blue) is broader than the Poisson posterior (orange), consistent with the single-realization result.
    }
    \label{fig:mean_inference}
\end{figure*}

Fig.~\ref{fig:mean_inference} shows the resulting correlation function fit and MCMC posterior. By construction, the data points coincide with the true model (black curves), and both the covariance and Poisson fits recover the input masses. However, the covariance-based posterior remains substantially broader than the Poisson posterior, confirming that the difference in uncertainty estimates is a systematic property of the error model rather than a consequence of noise in any individual realization.

\begin{figure*}
\centering
    \begin{minipage}{0.24\textwidth}
        \centering
        \textbf{$\log (\mming/\msun) = 10.5$} \\
        \includegraphics[width=\textwidth]{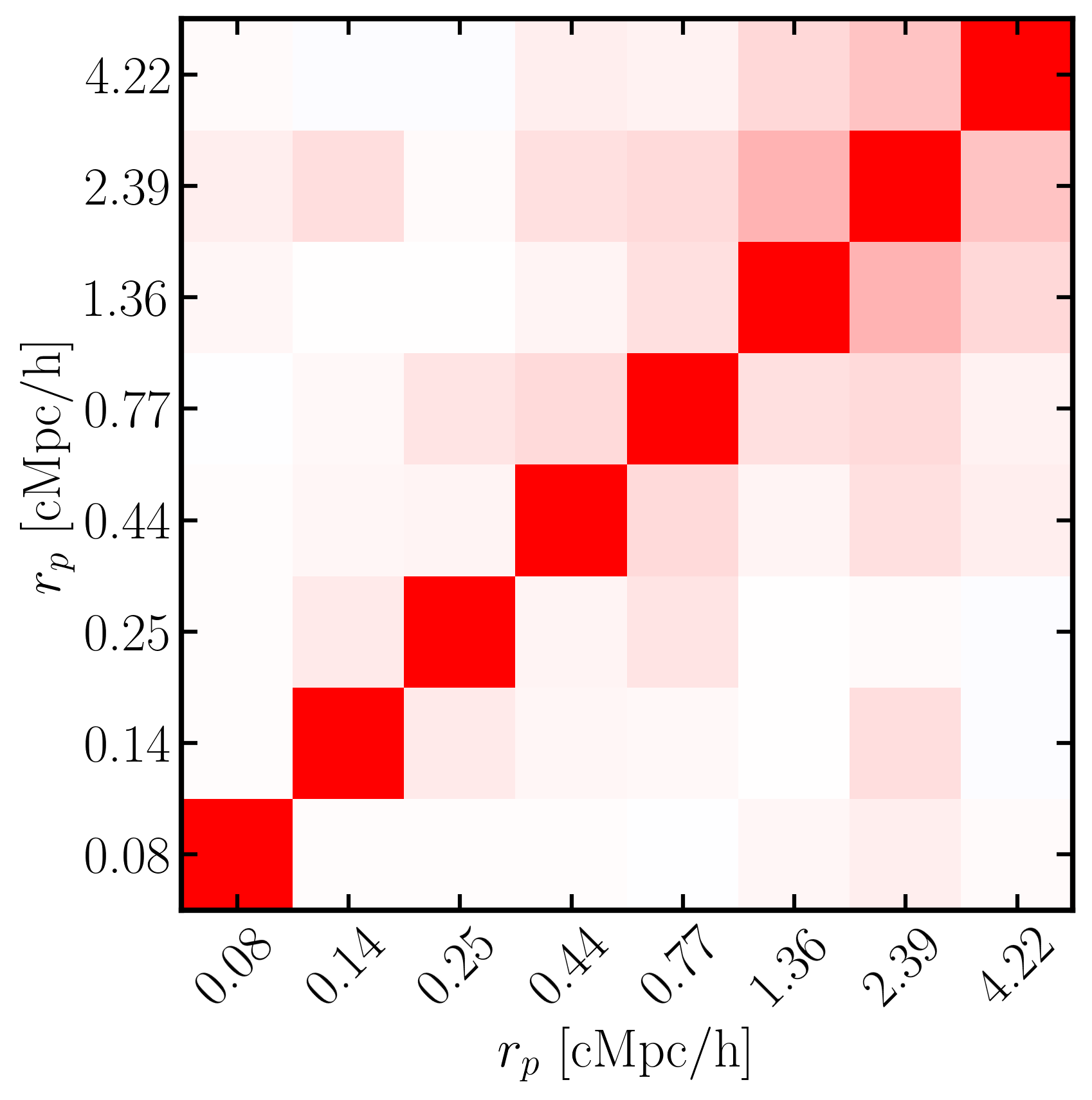}
    \end{minipage}%
    \begin{minipage}{0.24\textwidth}
        \centering
        \textbf{$\log (\mming/\msun) = 10.7$} \\
        \includegraphics[width=\textwidth]{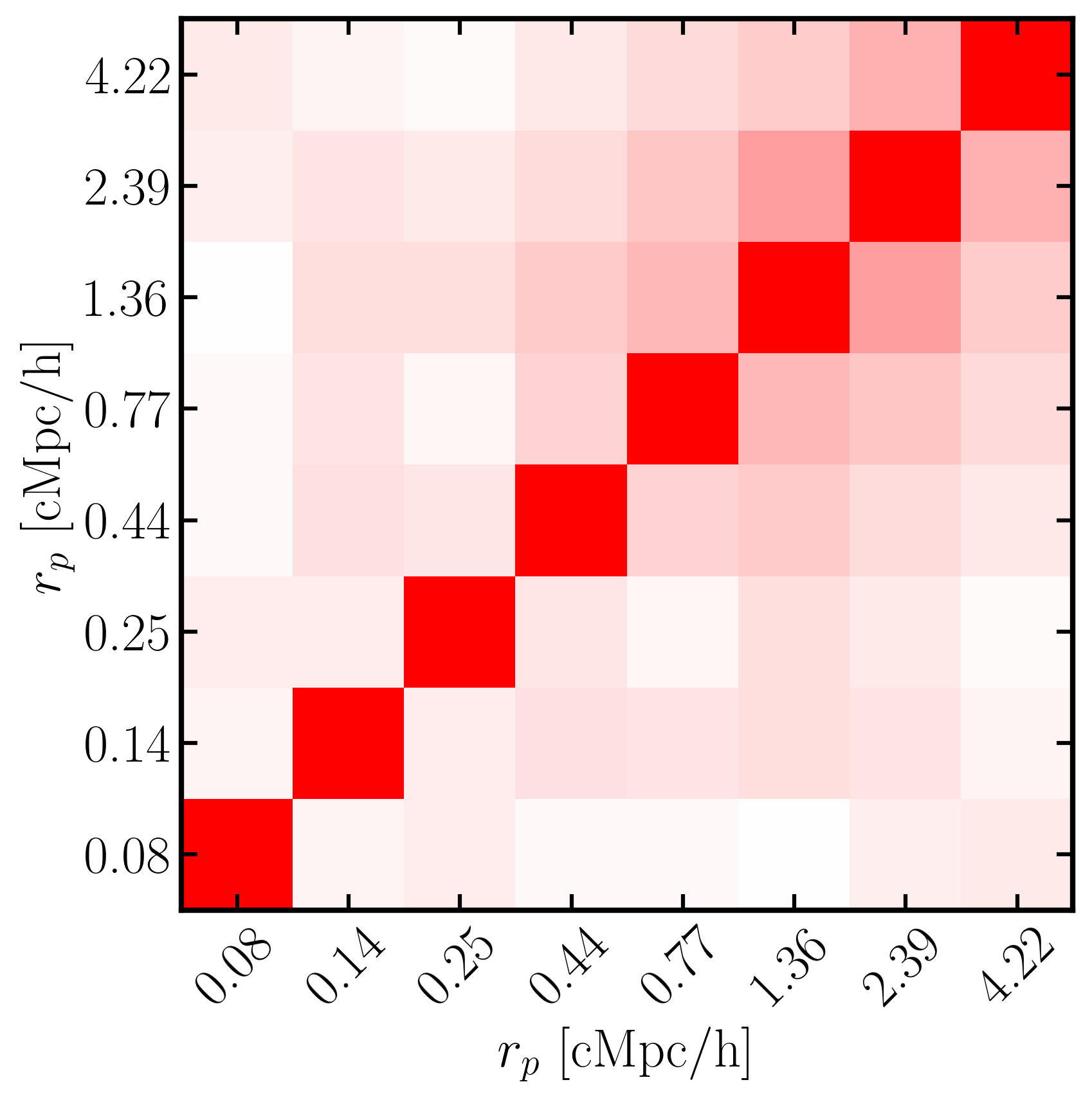}
    \end{minipage}%
    \begin{minipage}{0.24\textwidth}
        \centering
        \textbf{$\log (\mming/\msun) = 10.9$} \\
        \includegraphics[width=\textwidth]{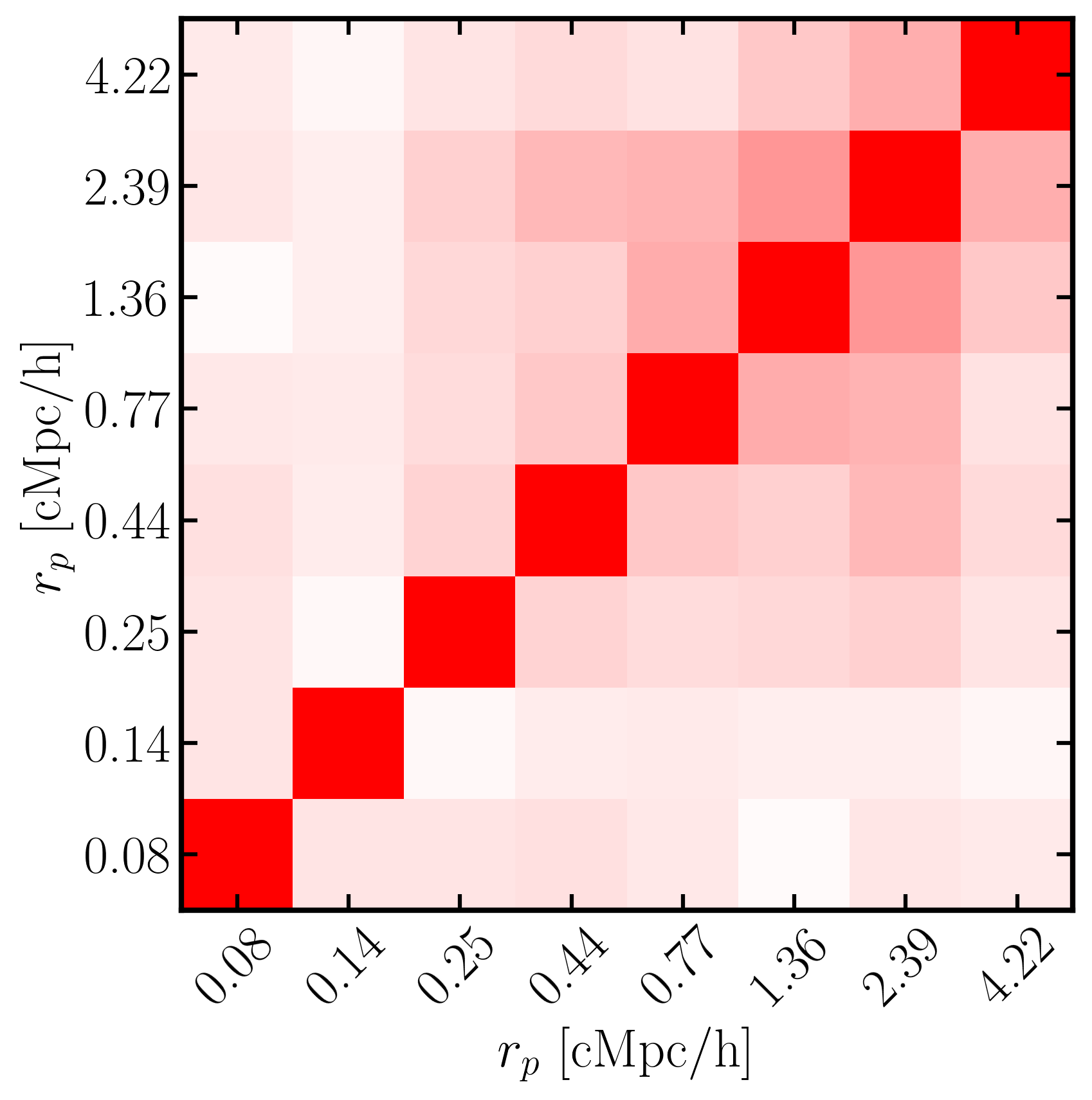}
    \end{minipage}%
    \begin{minipage}{0.24\textwidth}
        \centering
        \textbf{$\log (\mming/\msun) = 11.1$} \\
        \includegraphics[width=\textwidth]{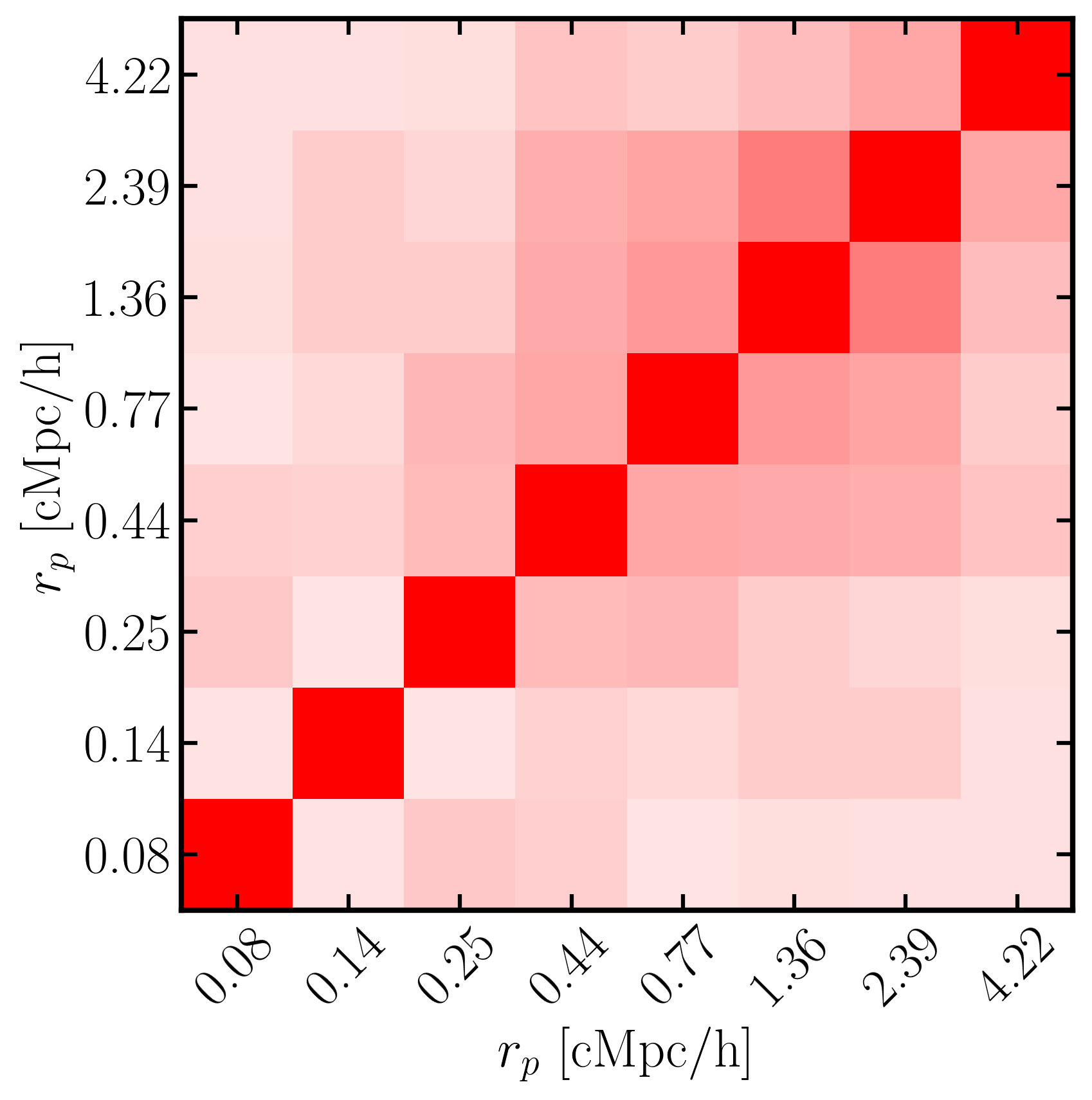}
    \end{minipage}

    \begin{minipage}{0.5\textwidth}
        \centering
        \includegraphics[width=\textwidth]{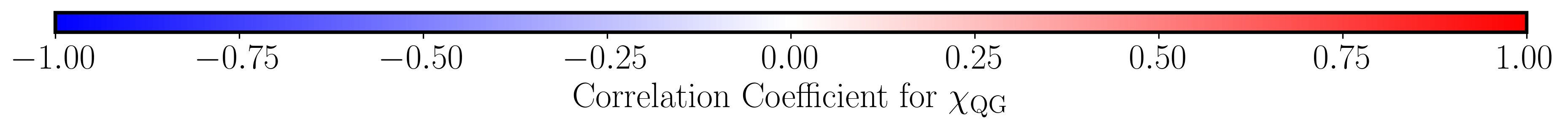}
    \end{minipage}

    \caption{Similar to Fig.~\ref{fig:cov_cross}, but showing how the correlation structure of $\chi_{\rm QG}$ depends on the galaxy halo mass. From left to right, $\log(\mming/\msun)$ increases from 10.5 to 11.1, while the quasar halo mass is fixed at $\log(\mminq/\msun) = 12.5$. The off-diagonal correlation becomes slightly stronger as $\mming$ increases. Compared with Fig.~\ref{fig:cov_cross}, the correlation structure of $\chi_{\rm QG}$ depends more strongly on the tracer galaxy halo mass than on the quasar halo mass.}
    \label{fig:cov_cross_gal}
\end{figure*}

While Fig.~\ref{fig:cov_cross} shows how the correlation structure of $\chi_{\rm QG}$ varies with the quasar halo mass $\mminq$ at fixed $\mming$, Fig.~\ref{fig:cov_cross_gal} shows the complementary dependence on $\mming$ at fixed $\mminq$.

\section{Correlation Length}
\label{appendix:corr_length}

Many high-redshift clustering studies infer halo masses by fitting a power-law correlation function to extract the correlation length $r_0$, converting $r_0$ into an effective large-scale bias, and then mapping this bias onto a characteristic halo mass using analytical prescriptions \citep[e.g.,][]{Shen2007, Arita2023, Eilers2024}. To demonstrate that the Poisson error underestimation is not limited to the minimum halo mass inference presented in the main text, we repeat the covariance versus Poisson comparison for the power-law correlation length, following the same approach as in \citet{Huang2026}.
We model the real-space
correlation function as a power law $\xi(r)=(r/r_0)^{-\gamma}$, where $r_0$ is the correlation length and $\gamma$ is the slope, fixed at $\gamma=2$. The model prediction for each radial bin is obtained by substituting $\xi(\rp, \pi) = \bigl(\sqrt{\rp^2 + \pi^2}/r_0\bigr)^{-\gamma}$ into Eq.~\ref{eq:volavg_corr}, giving $\chi_{V,\rm model}(r_{p,\rm min}, r_{p,\rm max}; r_0, \gamma)$.

We note that, unlike the minimum halo mass inference in \S\ref{sec:mmin} where the covariance matrix varies with the model parameters $(\mming, \mminq)$ as written in Eq.~\ref{eq:auto_likelihood}, the power-law correlation length $r_0$ is not a parameter of the mock catalog generation. Since our mock realizations are constructed from step-function HOD models parameterized by $(\mming, \mminq)$, the covariance matrix cannot easily be made $r_0$-dependent. 
We therefore adopt a \textit{fixed} covariance matrix evaluated at the fiducial model $(\log\mming/\msun, \log\mminq/\msun) = (10.6, 12.2)$ for the $r_0$ fit, as is common when the covariance cannot be directly linked to the fitted parameters \citep[see, e.g.,][]{Norberg2009}.

Assuming deterministic bias, where galaxies and quasars trace the same underlying dark matter distribution, the quasar auto-correlation function can be expressed as $\xi_{\rm QQ} = \xi_{\rm QG}^2 / \xi_{\rm GG}$ \citep{Croom2001, GarciaVergara2017, Eilers2024}. Following \citet{Eilers2024} and \citet{Huang2026}, we jointly fit the \OIII-emitter auto-correlation and the quasar--\OIII-emitter cross-correlation to simultaneously constrain the quasar auto-correlation length $r_{0}^{\rm QQ}$ and the galaxy auto-correlation length $r_{0}^{\rm GG}$. The total likelihood is:
\begin{align}
\mathcal{L}_{\rm tot}(r_0^{\rm QQ}, r_0^{\rm GG}) =
&\mathcal{N}\left(\chi_{\rm GG}^{\rm obs} \middle| \chi_{\rm GG}^{\rm model}(r_0^{\rm GG}),\, \boldsymbol{C}_{\rm GG}\right) \nonumber \\
\cdot\, &\mathcal{N}\left(\chi_{\rm QG}^{\rm obs} \middle| \chi_{\rm QG}^{\rm model}(r_0^{\rm QQ}, r_0^{\rm GG}),\, \boldsymbol{C}_{\rm QG}\right),
\end{align}
where $\boldsymbol{C}_{\rm GG}$ and $\boldsymbol{C}_{\rm QG}$ are evaluated at the fiducial masses.

\begin{figure}
\centering
	\includegraphics[width=\columnwidth]{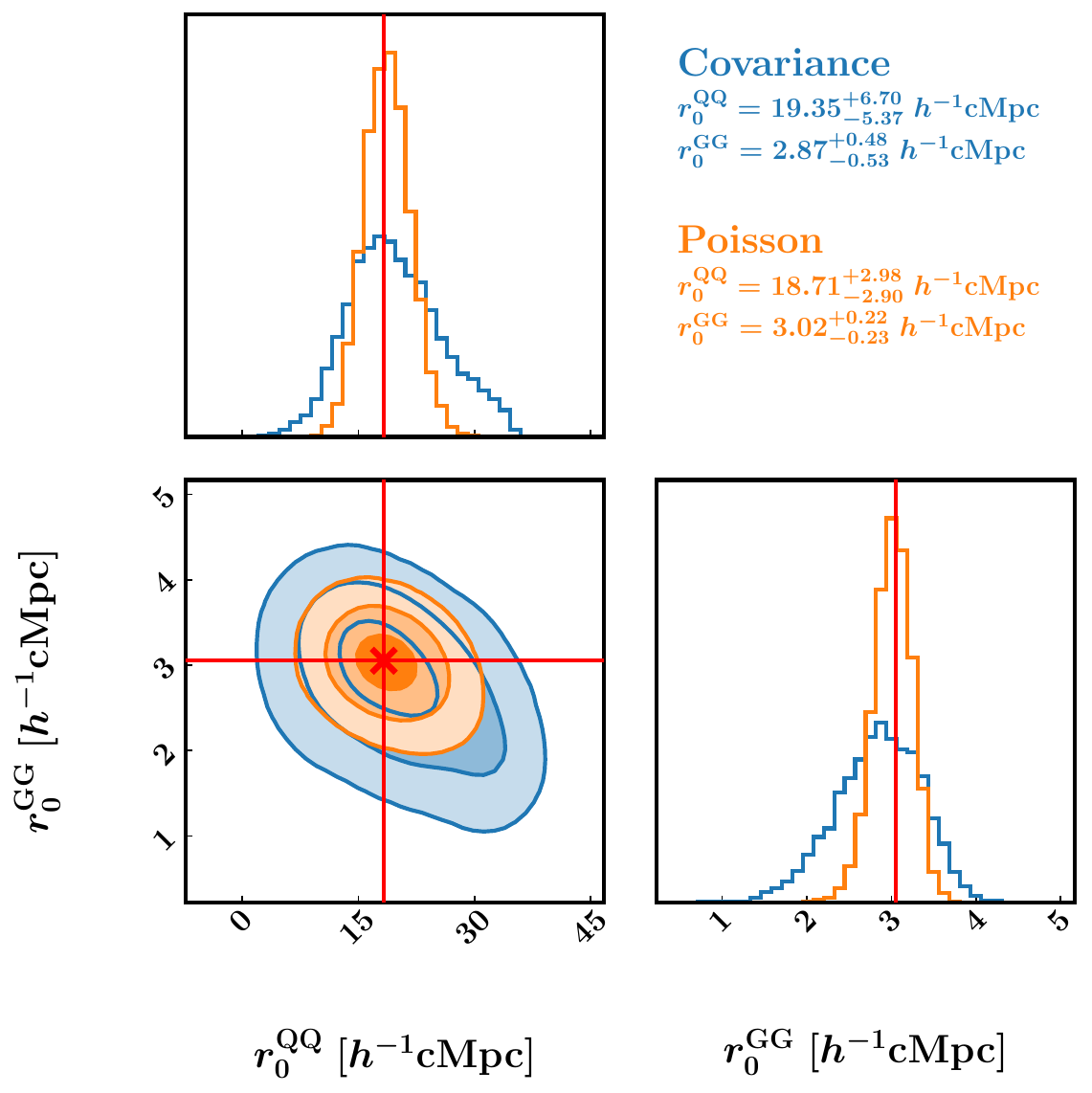}
    \caption{Joint posterior for the quasar auto-correlation length $r_{0}^{\rm QQ}$ and galaxy auto-correlation length $r_{0}^{\rm GG}$, from the joint power-law fit ($\gamma=2$) to $\chi_{\rm GG}$ and $\chi_{\rm QG}$ assuming deterministic bias ($\xi_{\rm QQ}=\xi_{\rm QG}^2/\xi_{\rm GG}$). The observed data is set to the true (noiseless) model. Blue filled contours use the full mock covariance; orange filled contours use Poisson errors. Red cross mark the true input values. The anti-correlation between $r_{0}^{\rm QQ}$ and $r_{0}^{\rm GG}$ arises from the coupling through the cross-correlation constraint. The Poisson posterior is systematically narrower, underestimating the true uncertainty.
    }
    \label{fig:corner_r0_true}
\end{figure}

Fig.~\ref{fig:corner_r0_true} shows the joint posterior for $(r_{0}^{\rm QQ}, r_{0}^{\rm GG})$ when the true power-law model is used as the observed data (i.e., noiseless). The anti-correlation between the two parameters is evident: the joint fit couples $r_{0}^{\rm QQ}$ and $r_{0}^{\rm GG}$ through the cross-correlation, so increasing $r_{0}^{\rm QQ}$ while decreasing $r_{0}^{\rm GG}$ can leave $\chi_{\rm QG}$ nearly unchanged. As with the minimum halo mass inference, the covariance-based posterior (blue) is broader than the Poisson posterior (orange), demonstrating that Poisson errors underestimate the uncertainty on the correlation lengths.

\bsp	
\label{lastpage}
\end{document}